\documentclass[]{aastex631}
\usepackage[figuresright]{rotating}

\usepackage{booktabs} 
\usepackage{subfigure}
\usepackage{graphics}
\shorttitle{PLZC relations for CBs}
\shortauthors{Song \& Tian}
\graphicspath{{./}{figures/}}

\begin{document}

\title{Period-Luminosity-Metallicity-Color Relations of Late-type Contact Binaries in the Big Data Era}

\correspondingauthor{Zhi-Jia TIAN}
\email{tianzhijia@ynu.edu.cn}

\author{Lian-Yun Song}
\affiliation{Department of Astronomy, Yunnan University,Kunming 650200,China}

\author[0000-0003-0220-7112]{Zhi-Jia Tian }
\affiliation{Department of Astronomy, Key Laboratory of Astroparticle Physics of Yunnan Province, Yunnan University, Kunming 650200, China}



\begin{abstract}

Binary stars ubiquitous throughout the universe are important. 
Contact binaries (CBs) possessing Period-Luminosity (PL) relations could be 
adopted as distance tracers.  
The PL relations of CBs are influenced by metallicity abundance and color index, which are connected to both the radius and luminosity of stars.
Here we propose fine relations of Period-Luminosity-Metallicity-Color (PLZC) 
from the ultraviolet to infrared bands based on current surveys. 
The accuracy of the distance estimation is 6\% and 8\%, respectively, depending on the PLZC relations of the CBs in the infrared and optical bands of the collected data. 
PLZC models are still more accurate than PLC models in determining intrinsic luminosity, notwithstanding their limited improvement.
Meanwhile, these relations based on synthetic photometry are also calibrated.
On the basis of the synthetic photometry, a 6\% accuracy of distance is estimated. 
The measured or synthetic data of PLZC or PLC relations in infrared bands comes first in the list of suggestions for distance estimations, and is followed by the measured data of optical bands.
\end{abstract}


\section{Introduction} \label{sec:intro}
In the Milky Way, more than half of stars belong to binary systems, 
which are important to study in many fields, 
such as stellar physics \citep[][]{2023A&ARv..31....1A}, 
Galaxy and extra-galaxies \citep{2009ApJS..182..216K}, 
cosmology \citep[][]{2013Natur.495...76P}, 
and gravitational wave astronomy \citep[][]{2017ApJ...848L..12A}. 
When the host star is filled with Roche critical volume, the process of 
material exchange begins accompany with the essence of energy exchange. 
Therefore, material and energy exchanges are inseparable physical processes,
which provide conditions for the formation of some special stars, 
such as
red novae \citep{THK+2011,MM+2022}, type Ia supernovae \citep{WR+1984,IT+1984,WI+1973}, short gamma-ray bursts \citep{SMT+2006,FE+2013},
binary black hole mergers \citep{A+2016}, and kilonovae \citep{SC+2017,CB+2017}.
        
These binary systems with significant material exchange between the 
primary and its companion star are called close binary stars, 
which are common among binary systems. 
According to the overall evolution, close binary stars are divided into 
three types: detached, semi-detached, and contact phases. 
Contact binary stars (CBs) with two companions full of their own Roche lobes 
share a common external envelope. Then tidal interactions make them rotate 
rapidly and produce magnetic activities similar to the Sun \citep{DJP+2020,LHK+2004,K+2011}.
Generally, binary systems with common envelope will lose the shell through the 
outer Lagrange point, thus becoming a system with very short orbital period. 
Most importantly, CBs can be used as distance indicators based on 
Period-Luminosity relation \citep{1994PASP..106..462R}. 

Since CBs proposed as distance tracers by \citet{E+1967}, various studies have
tried to establish Period-Luminosity (PL) and Period-Luminosity-Color (PLC) relations of CBs.
This issue has also attracted attention and research of \citet{R+1974}. 
\citet{M+1981} derived the PL relation based on a sample of 37 CBs, 
but the work mainly focuses on the mass-brightness relation. Because the effective temperature and luminosity are interrelated, 
it is necessary to consider the color index term for calibrating the absolute magnitude. 
\citet{1994PASP..106..462R} calibrated the absolute magnitude based on $B$, $V$,
and $I$-band of CBs in open clusters,
established the first accurate PLC relations. The relations are re-calibrated 
based on 40 CBs with Hipparcos parallax \citep{RD+1997}.
Then, \citet{EBY+2009} analyzed 31 CBs with the most accurate distance by 
applying Lutz-Kelker corrections \citep{LK+1973} to the parallax from the 
Hipparcos catalog, and re-calibrated the classic PLC relation of \citet{RD+1997}. 
Meanwhile, they extended the PLC relations to the near-infrared $J$, $H$, and $K_{s}$-band. 
Independently, \citet{CDD+2016} calibrated the distances of a sample of 66 CBs 
from open clusters based on Hipparcos parallaxes to derived near-infrared 
$J$, $H$, and $K_{s}$-band PL relations. 
Furthermore, \citet{CDD+2018} established PL relations for 12 bands 
from optical to mid-infrared bands based on 183 nearby CBs (within 300 pc) with 
accurate Tycho-Gaia parallax \citep{G+2016}, obtaining that the minimum scatter of 
absolute magnitude is 0.16 mag. 
For the first time, PL and PLC relations for late-type CBs are established 
by \citet{NLB+2021} based on $g$, $r$, and $i$-band light curves collected by Zwicky 
Transient Factory \citep[ZTF,][]{GMK+2019}. 

In earlier work of \citet[][]{R+1995,R+2000}, they discussed that metallicity may 
have an impact on the PL relation.
The PL relation were more susceptible to redness correction uncertainties, so it was
not necessary to retain the metallicity term in the PL relation \citep{R+2004}. 
However, CBs will become brighter about 0.2 mag in the near-infrared bands when the metallicity ($[\rm Fe/H]$) decreases by 1 dex \citep{CDD+2016}. 
Benefited from the coverage and precision of current surveys, 
more accurate PL relations for CBs in a large sample of all sky could be established.
 
In this work, we makes comparisons between PL, Period-Luminosity-Metallicity (PLZ), PLC, and Period-Luminosity-Metallicity-Color (PLZC) relations based on currently surveys.
The dispersion of PLZC relation is minimal for ultraviolet to infrared bands.
In Sect. \ref{sec:data}, we describe the data of CBs adopted in this work. 
In Sect. \ref{sec:calibration}, the fitting models and results are presented, 
followed by discussions and conclusions in Sect. \ref{sec:dis_con}.

\section{Data} \label{sec:data}
With the increase of high-cadence, long-term, and high-precision photometric surveys 
in recent years, more and more CBs have been identified, which construct a large 
sample to carry out various of researches.   
The data of CBs analyzed in this work are collected based on the General Catalogue of 
Variable Stars \citep[Version GCVS 5.1,][]{SKD+2017}, 
the All-Sky Automated Survey for Supernovae \citep[ASAS-SN,][]{JSKSK+2019}, 
and the Northern Catalina Sky Survey \citep[CSS,][]{SDD+2020}. 
The latest released data from those catalogues contains 3261, 76378, and 2335 CBs, respectively.

The photometric data analyzed in this work covers the bands of Gaia DR3 \citep{2022yCat.1355....0G}, 
WISE \citep{WISE+2021} and GALEX \citep{BS+2017}.
The distances (parallaxes) of the CBs are adopted from Gaia DR3. 
The LAMOST spectroscopic survey provides the largest database of low and median-resolution 
spectra to measure stellar atmospheric parameters and radial velocities for millions of stars \citep{2012RAA....12.1197C,2012RAA....12..723Z,2012RAA....12..735D,2014IAUS..298..310L,2015MNRAS.448..855Y}. 
Metallicity abundance ($[\rm Fe/H]$) and effective temperature ($T_{\rm eff}$) of the CBs are obtained through 
cross-matching with the catalogs of LAMOST Data Release 9\footnote{The LAMOST DR9 catalogs are available via \url{http://www.lamost.org/dr9/v1.0}.}, 
which are derived through the LAMOST stellar parameter pipeline \citep[LASP,][]{L+2015}. 
In total, 6883 unique CBs are selected with complete information such as photometric, astrometric, and spectroscopic parameters. 
A summary of the CBs is shown in Table \ref{table1}. The duplicate data has been removed to optimize redundancy, and the total number of duplicates is shown in parentheses.
\begin{deluxetable}{cccccch}
\tablecaption{CBs collected from different surveys or catalogs. \label{table1}}
\tablewidth{0pt}
\setlength{\tabcolsep}{5mm}
\tablehead{
Surveys & Number of CBs &  LAMOST DR9\tablenotemark{a} & Gaia DR3\tablenotemark{b} & WISE\tablenotemark{c} & GALEX\tablenotemark{d}}
\startdata
GCVS    & 3,261   & 680    & 655   & 652  & 368  \\
ASAS-SN & 76,378  & 6,503  & 6,221 & 6,110   & 2,447   \\
CSS     & 2,335   & 720    & 704   & 694   & 359   \\
Total   & 81,974  & 7,903  & 6,883(697) & 6,772(685) & 2,844(330) \\
\enddata
\tablenotetext{a}{Number of CBs after further cross-match with LAMOST DR9.}
\tablenotetext{b}{Number of CBs after further cross-match with Gaia DR3.}
\tablenotetext{c}{Number of CBs for calibration covered by WISE.} 
\tablenotetext{d}{Number of CBs for calibration covered by GALEX.} 
\end{deluxetable}

\begin{figure}[ht!]
\plotone{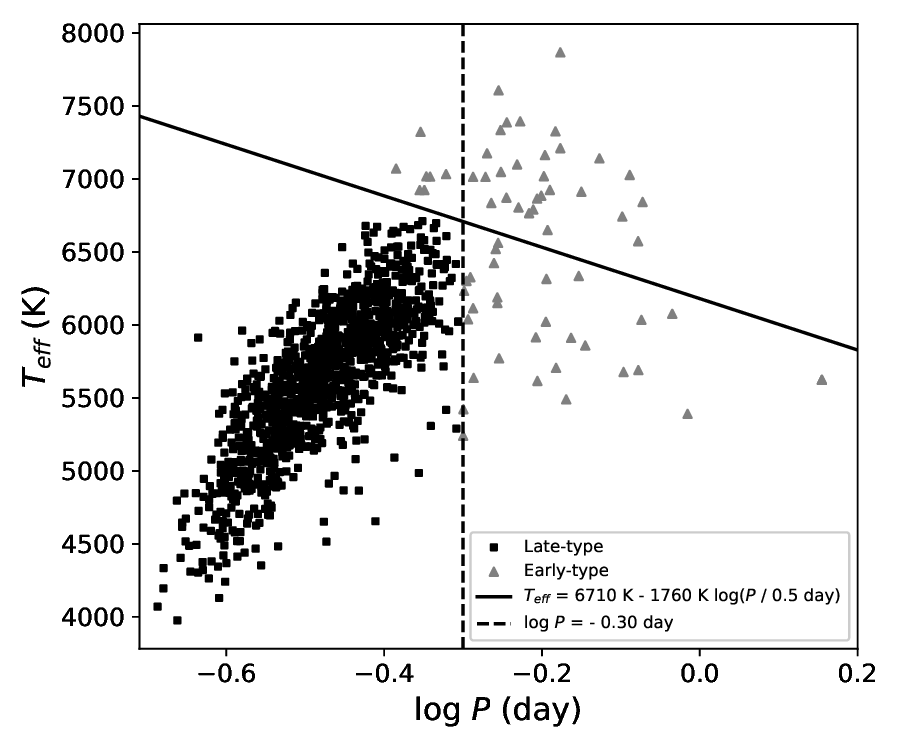}
\caption{$\rm T_{eff}$ versus $\log P$ for CBs in our sample.
Black squares denote the late-type CBs, while gray triangles early-type ones.
\label{fig1}}
\end{figure}

Based on the different distributions of orbital periods and spectral characteristics between early and late-type CBs, 
late-type ones are selected by using a period-temperature relation or a period cut.
We apply the both methods to limit the sample to avoid mixing early and late-type CBs as much as possible.
Figure \ref{fig1} presents the $T_{\rm eff}$ versus $\log P$ of the CBs.  
The solid and dashed lines present the thresholds, $T_{\rm eff} = 6710 \ \rm K -1760 \ K \log (P / 0.5 \ day)$ and $\log P = -0.30$ day given in \citet{2020MNRAS.493.4045J}, respectively. 
After applying the thresholds, the black squares are classified into late-type CBs, while the gray triangles are considered to be the early-type ones.
For the remaining late-type CBs, 
to make sure the reliability of the following calibrations, we exclude the inaccurate distance measurement data with $\tt Parallax\_over\_error$ less than 80.
We also exclude some sources with large absolute magnitude errors, 
such as Gaia $G$, $B_{p}$, and $R_{p}$-band, whose absolute magnitude errors are larger than 0.04, 0.07, and 0.07 mag, respectively. 
Therefore, the final sample of CBs for calibrations are listed in Table \ref{gaiatable}, which could be download online.
\begin{deluxetable}{lllllllh}
\caption{The final sample of CBs for calibrations.} 
\label{gaiatable}
\tablewidth{0pt}
\setlength{\tabcolsep}{3.5mm}
\tablehead{
Clo. & Name & Units & Description &  Example  &   }
\startdata
1    & Gaia\_source\_id & ---   & Gaia SOURCE\_ID           & 1038711452258591744      &...  \\
2    & RA               & deg   & Degree of Right Ascension (J2000)  & 142.236     &...\\
3    & Dec              & deg  & Degree of Declination (J2000)      & 59.935     &...\\
4    & Parallax         &mas   & Gaia DR3 parallax                  & 5.599         &...   \\
5    & Parallax\_over\_error & --- & Parallax divided by its standard error                            & 450.226    &...  \\
6    & Period           &day   & Orbital period                     & 0.2488    &... \\
7    & $T_{\rm eff}$    &K   & Effective temperature from LAMOST  &4851     &... \\
8    & $T_{\rm eff}$\_err & K  & Effective temperature  error               & 32      &... \\
9    & [Fe/H]           &dex   &[Fe/H] from LAMOST                  & -0.067      &...  \\
10   & [Fe/H]\_err       &dex  &  [Fe/H] error       & 0.027        &...  \\
11   & $\langle FUV\rangle$ &mag  & GALEX FUV\_band mean magnitude  & 21.806      &... \\
12     &  FUV\_err          &mag &  Error on FUV\_band mean magnitude   & 0.231      &... \\
13   & $\langle NUV\rangle$ &mag  & GALEX NUV\_band mean magnitude  & 18.973     &...  \\
14     & NUV\_err           &mag &  Error on NUV\_band mean magnitude    & 0.023     &...    \\
15   & $\langle Bp\rangle$  &mag  & Gaia DR3 Bp\_band mean magnitude&12.537    &...  \\
16     & Bp\_err            &mag &  Error on Bp\_band mean magnitude     &0.010   &...  \\
17   & $\langle G\rangle$   &mag  & Gaia DR3 G\_band mean magnitude & 12.025    &...  \\
18   & G\_err             &mag &  Error on G\_band mean magnitude                                  & 0.004     &... \\
19   & $\langle Rp\rangle$  &mag  & Gaia DR3 Rp\_band mean magnitude& 11.336  &... \\
20   & Rp\_err            &mag &  Error on Rp\_band mean magnitude                                  & 0.008    &... \\
21   & $\langle W1\rangle$  &mag  & WISE W1\_band mean magnitude    & 9.799       &...  \\
22   & W1\_err            &mag &  Error on W1\_band mean magnitude                                 & 0.022          &... \\
23  & $\langle W2\rangle$  &mag  & WISE W2\_band mean magnitude    & 9.831       &...  \\
24   & W2\_err            &mag &  Error on W2\_band mean magnitude                                  & 0.02        &... \\
25   & $\langle W3\rangle$  &mag  & WISE W3\_band mean magnitude    & 9.787        &... \\
26   & W3\_err            &mag &   Error on W3\_band mean magnitude                                 & 0.034      &... \\
27   & $\langle W4\rangle$  &mag  & WISE W4\_band mean magnitude    & 8.846        &... \\
 28 & W4\_err           &mag  &   Error on W4\_band mean magnitude                                 & 0.331       &... \\
\enddata
\tablecomments{The Table is published in its entirety in the machine-readable format.
A portion is shown here for guidance regarding its form and content.}
\end{deluxetable}

The extinction is estimated through the full-sky 3D dust-reddening map\footnote{\url{http://argonaut.skymaps.info/}} of $E(B-V)$ 
based on the dust thermal emission generated by \citet[]{Green_2019}. 
A value of 3.1 for $R_{\rm v}$ is adopted to estimate the extinction by applying 
the equation $A_{\rm v} = R_{\rm v} E(B-V)$. Additionally, extinction values in other bands are estimated based on the relative extinction values $A_{\lambda}/A_{\rm v}$ from \citet{J+2010}, \citet{W+2018}, \citet{2019ApJ...877..116W}, and \citet{RLM+1985}. The absolute magnitudes of the sample are deduced through the following equation, 
\begin{equation}
\label{amag}
M= m - 5\log D + 5 - A_{\rm \lambda}, 
\end{equation}
where $M$ and $m$ are absolute and apparent magnitudes, respectively. The quantity $D$ denotes the distance from Gaia astrometry, and $A_{\rm \lambda}$ is extinction described previously.

\begin{figure*}[htb]
\gridline{\fig{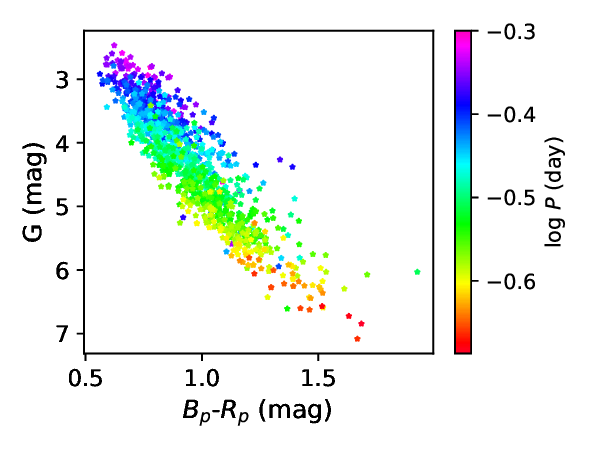}{0.5\textwidth}{}
          \fig{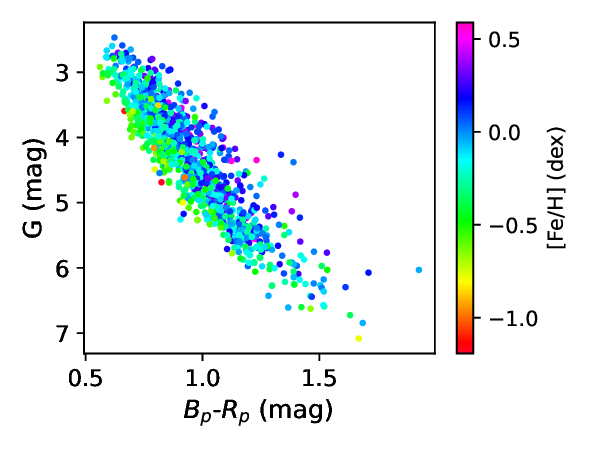}{0.5\textwidth}{}}
\caption{CBs distribution on CMD. G and $B_{p} - R_{p}$ denote the extinction corrected absolute
magnitude and color index, respectively. The colorbars in left and right panels denote 
Period ($\log P$) and metallicity abundance ($[\rm Fe/H]$), respectively. \label{fig2}}
\end{figure*}

The left panel of Fig. \ref{fig2} shows the locations of the CBs shift with periods 
in color-magnitude diagram (CMD). It demonstrates that the luminosity of late-type 
CBs are related to periods as well as stellar colors (e.g. $B_{p} - R_{p}$ color index from Gaia). 
Meanwhile, the right panel of the figure presents that metallicity-poor stars are bluer than metallicity-rich ones, 
which demonstrates the influences of metallicity on the positions of CBs in CMD. 
Therefore, the positions of CBs on CMD are related both to the period and $[\rm Fe/H]$, 
which confirms the influences of metallicity and color index on PL relations.

\section{Calibration of PL relations for late-type CBs} \label{sec:calibration}
The PLZC relations are derived with the periods, mean magnitude, metallicity abundance and color index of $B_{p}-R_{p}$ described in Sect. \ref{sec:data}. 
As comparisons, the PL, PLZ, and PLC relations are derived simultaneously. 
These relations are calibrated by fitting the following equations, 
\begin{equation} 
\label{equ1}
M = a \log P+b,
\end{equation}
\begin{equation}
\label{equ2}
M=a\log P +b+ c [\rm Fe/H], 
\end{equation}
\begin{equation}
\label{equ3}
M=a\log P +b+ d ( B_{p}-R_{p}), 
\end{equation}
\begin{equation}
\label{equ4}
M=a\log P +b+ c{\rm[Fe/H]} + d(B_{p}-R_{p}), 
\end{equation}
where $a$, $b$, $c$, and $d$ are set as free parameters to fit. 
To obtain the best-fitting relations, we refer to a {\tt Python} implementation, {\tt emcee}, of Markov Chain Monte Carlo (MCMC) proposed by \citet{GW+2010}. 
\subsection{PLZC and other relations for measured data} \label{uv}
\begin{figure*}
\gridline{\fig{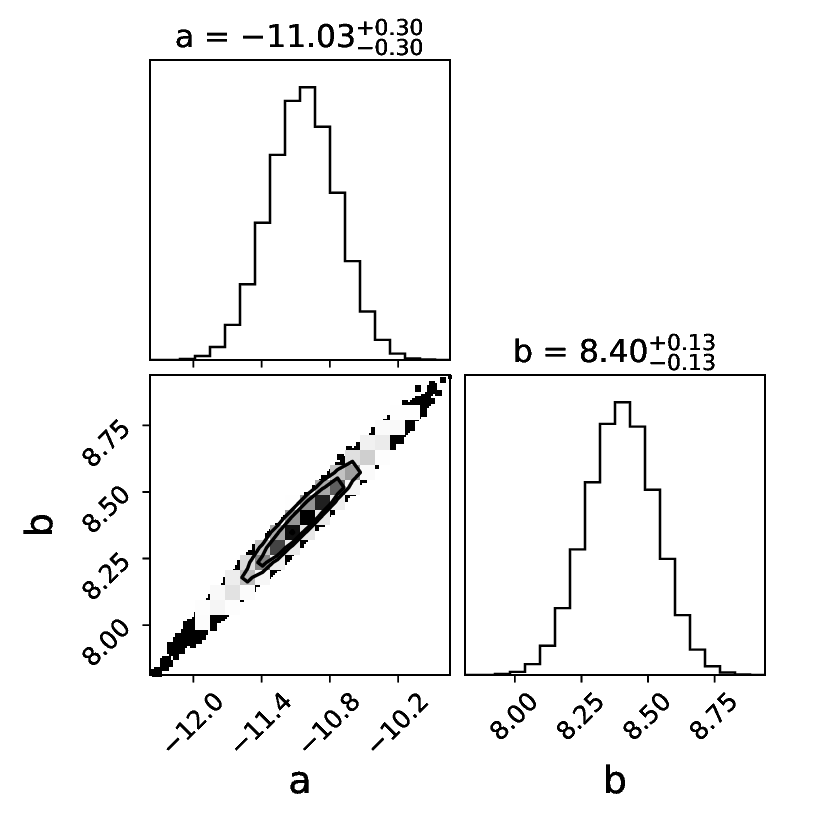}{0.34\textwidth}{(1) FUV-band}\hspace{-15mm}
         \fig{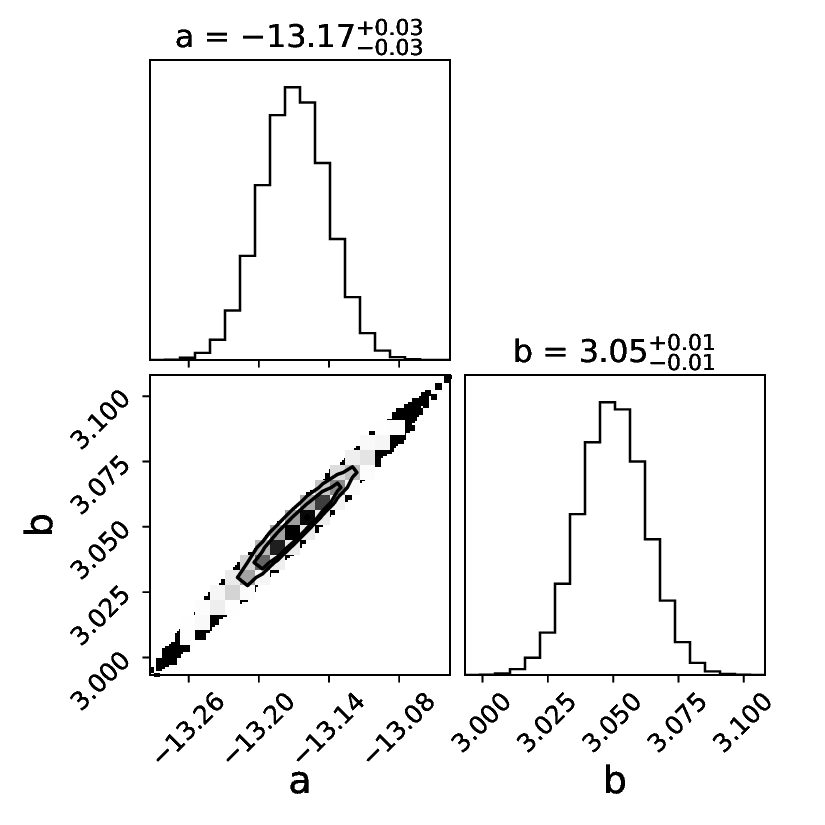}{0.34\textwidth}{(2) NUV-band}\hspace{-15mm}
          \fig{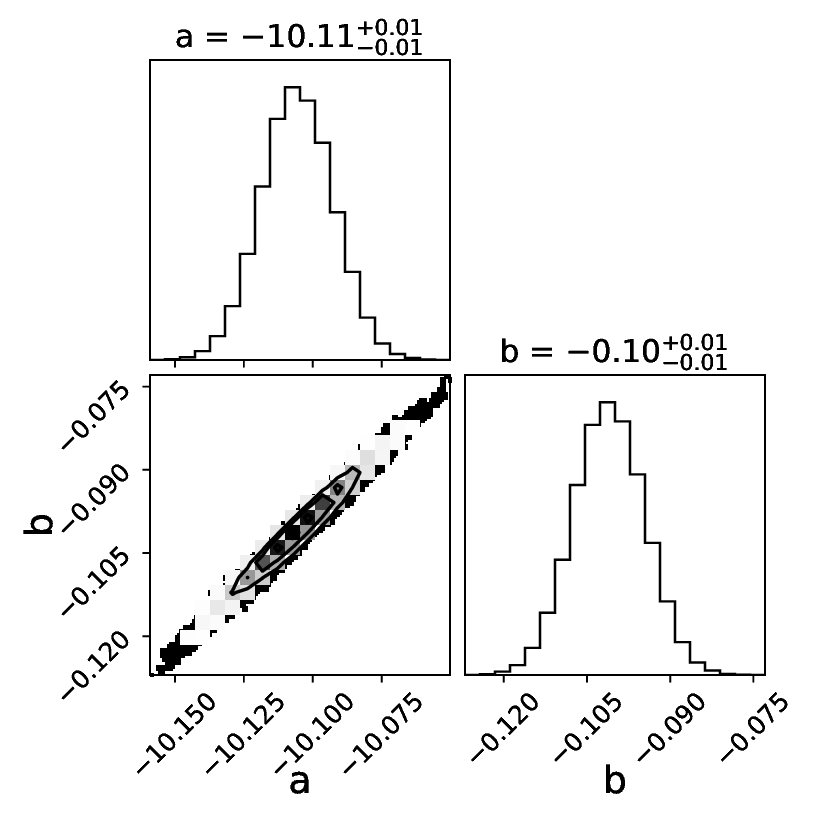}{0.34\textwidth}{(3) $B_{p}$-band}}
 \gridline{\fig{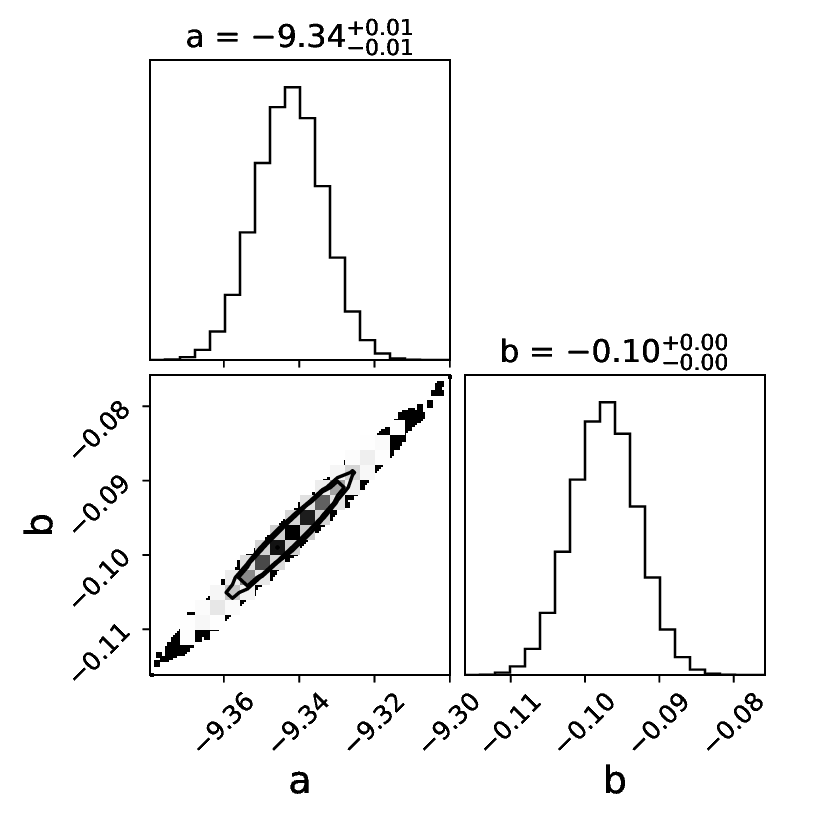}{0.34\textwidth}{(4) G-band}\hspace{-15mm}
           \fig{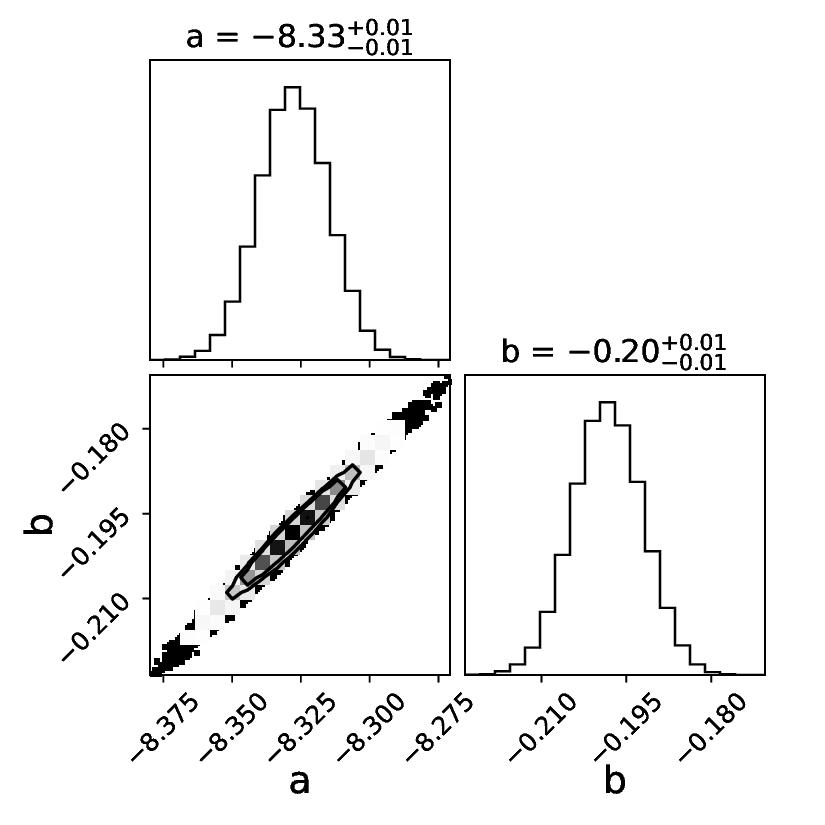}{0.34\textwidth}{(5) $R_{p}$-band}\hspace{-15mm}
          \fig{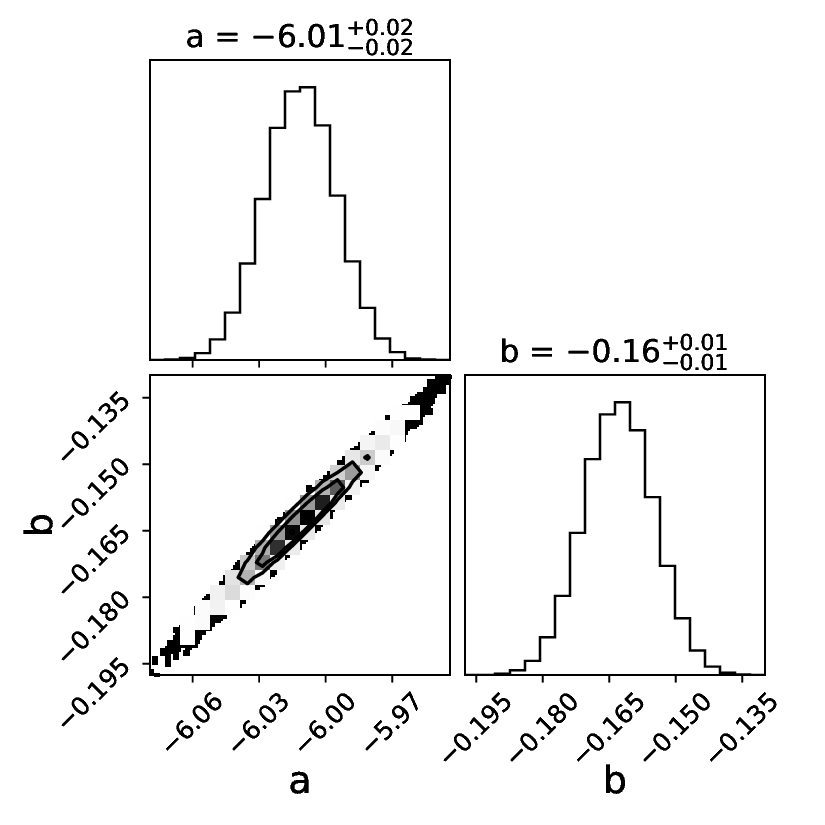}{0.34\textwidth}{(6) W1-band}}    
\gridline{\fig{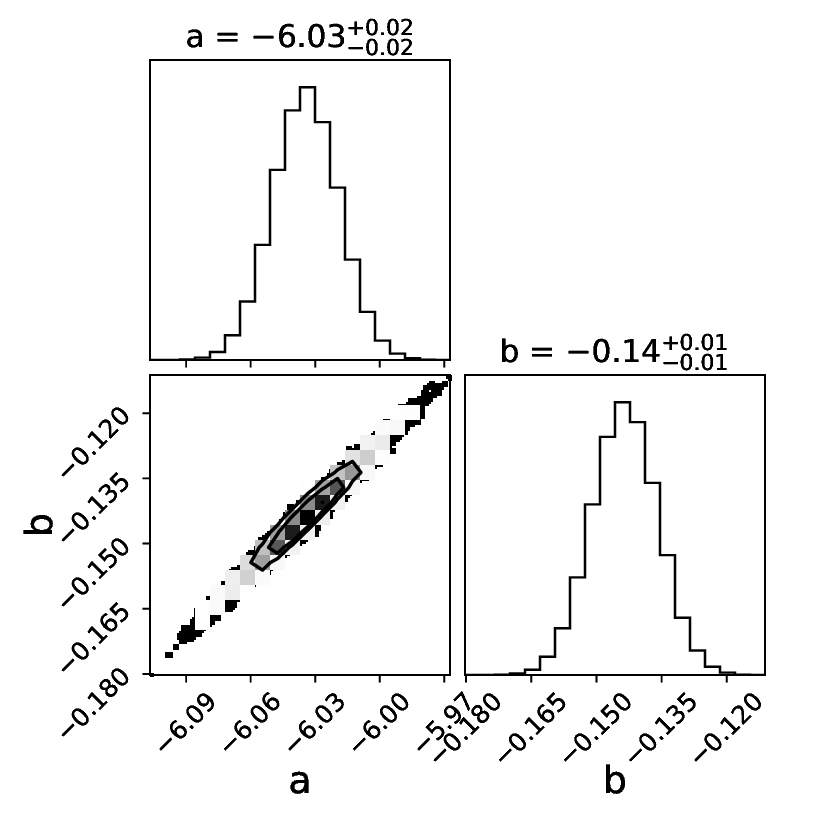}{0.34\textwidth}{(7) W2-band}\hspace{-15mm}
          \fig{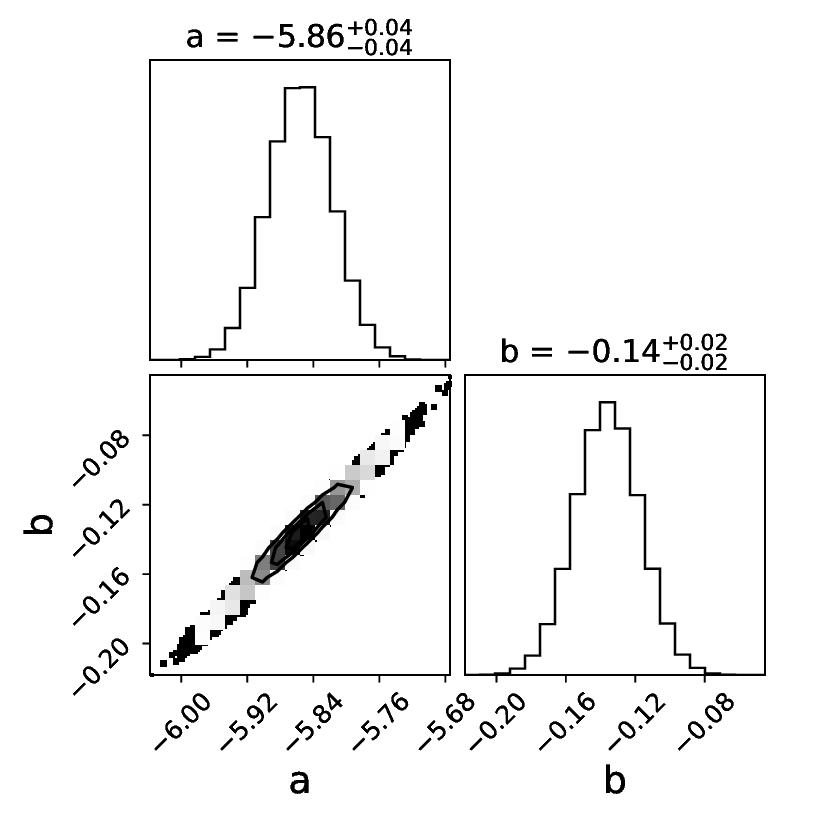}{0.34\textwidth}{(8) W3-band}\hspace{-15mm}
          \fig{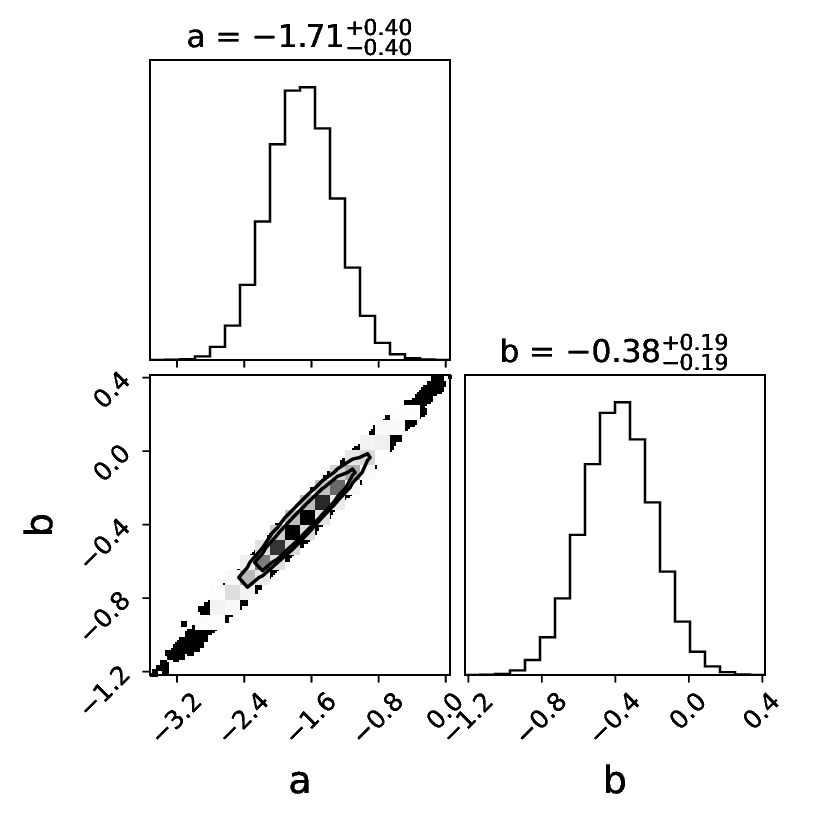}{0.34\textwidth}{(9) W4-band}}
\caption{Parameters fitting for PL relations by {\tt emcee} in each band, and tops of the panels are the best parameters obtained by fitting.
\label{fig.3}}
\end{figure*}

\begin{figure}
\gridline{\fig{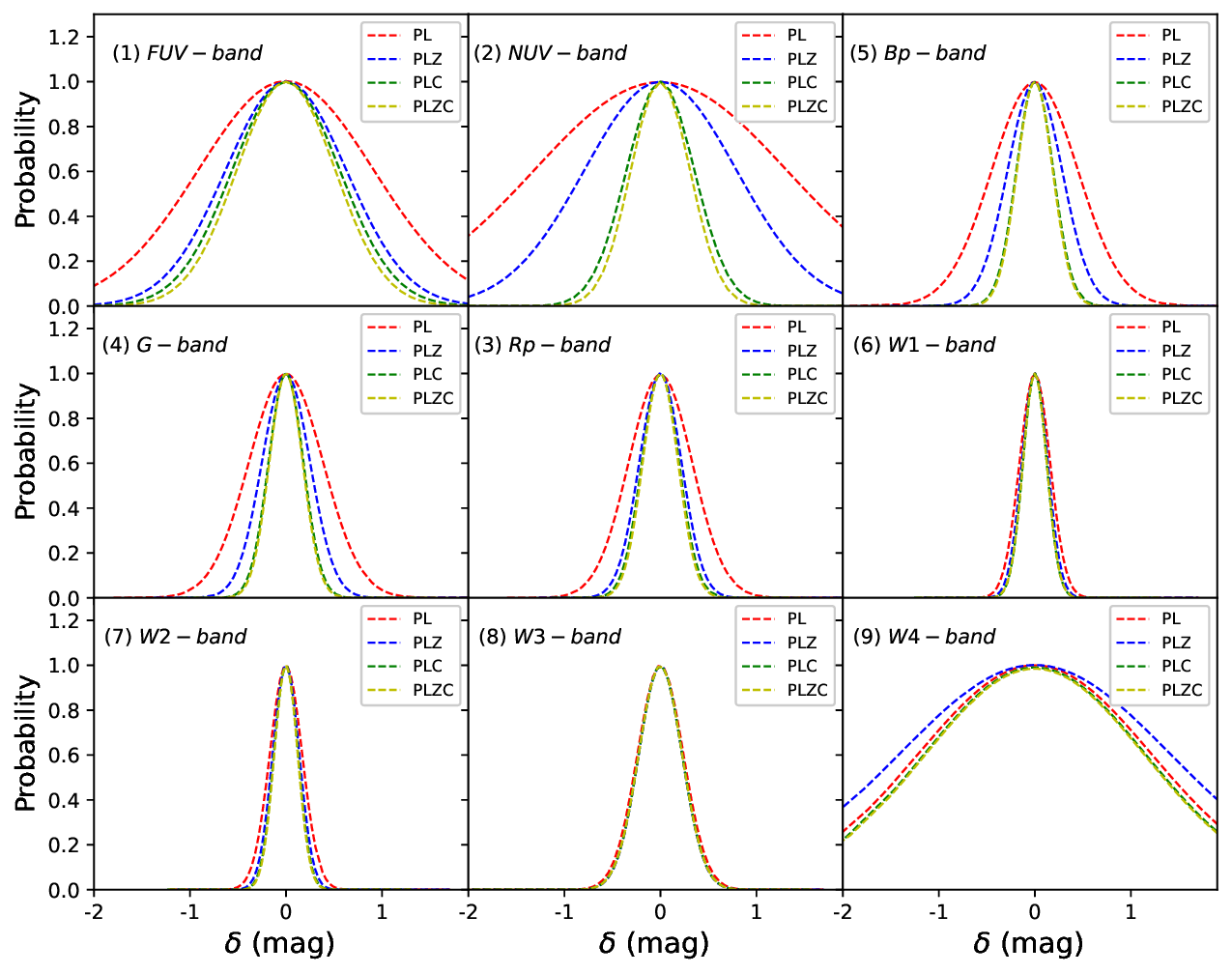}{0.9\textwidth}{}}    
\caption{Distributions of $\delta$ for each bands. The fitted Gaussian function to the histogram is shown as dash line.
\label{fig.4}}
\end{figure}

The PL relations for each band are calibrated firstly.
The fitting performances of the PL relations from ultraviolet to infrared bands are shown as corner plots in Fig. \ref{fig.3}.  
These corner plots show that the PL models are well fitted.
The coefficients and their errors of the best-fitting relations are texted at the tops of the panels. 
The numbers used for determining these relations are in the second column of Table \ref{table2}. 
Their slopes and intercepts are listed in the third and fourth columns.
To quantify the differences between absolute magnitudes from observations and these from models, 
a quantity $\delta$ is defined as,
\begin{equation}
\delta = M_{obs}-M_{model},
\end{equation}
where $M_{obs}$ denote the absolute magnitude directly from observations, 
while $M_{model}$ is the absolute magnitude deduced through equations 
\ref{equ1} - \ref{equ4} based on period and other observed quantities.
Then, the dispersion, $\sigma$, of the calibrated relations are derived by fitting the $\delta$ distribution with a Gaussian profile. 
The Gaussian-fitted distribution of $\delta$ are shown in Fig. \ref {fig.4}, represented by a red dotted line. 
The figure shows that the dispersion of PL relations in the FUV, NUV, and W4 bands is large. The large dispersion may be due to significant photometric errors and limited sample size \citep{2023ApJS..264...14Z}.
\begin{deluxetable}{ccccc}
\setlength{\tabcolsep}{8mm}
\tablecaption{Parameters of PL models for multi-band. \label{table2}}
\tablewidth{0pt}
\tablehead{
Band & Number &  a & b & $\sigma$ (mag) }
\startdata
$FUV$ & 199  & $-$11.03($\pm0.3$)  & 8.40($\pm0.13$) & 1.17 \\
$NUV$ & 632 & $-$13.17($\pm0.03$)  & 3.05($\pm0.01$) & 1.45 \\
\hline
$B_{p}$  & 1,093  & $-$10.11($\pm0.01$)  & $-$0.10($\pm0.01$) & 0.45 \\
$G$  & 1,093 & $-$9.34($\pm0.01$)  & $-$0.10($\pm0.01$)  & 0.37  \\
$R_{p}$  & 1,093 & $-$8.33($\pm0.01$)  & $-$0.20($\pm0.01$) & 0.33 \\
\hline
$W1$  & 1,019  & $-$6.01($\pm0.20$)  & $-$0.16($\pm0.01$) & 0.16 \\
$W2$  & 1,019  & $-$6.03($\pm0.20$) & $-$0.14($\pm0.01$)  & 0.17  \\
$W3$  & 992 &$-$5.86($\pm0.04$)  & $-$0.14($\pm0.02$) & 0.24 \\
$W4$  & 127& $-$1.71($\pm0.19$) & $-$0.38($\pm0.01$)  & 1.21  \\
\enddata
\tablecomments{The parameters a and b are the best-fitting coefficients in equation \ref{equ1}, and $\sigma$ are the dispersion of PL relations.}
\end{deluxetable}

\begin{figure*}
\gridline{\fig{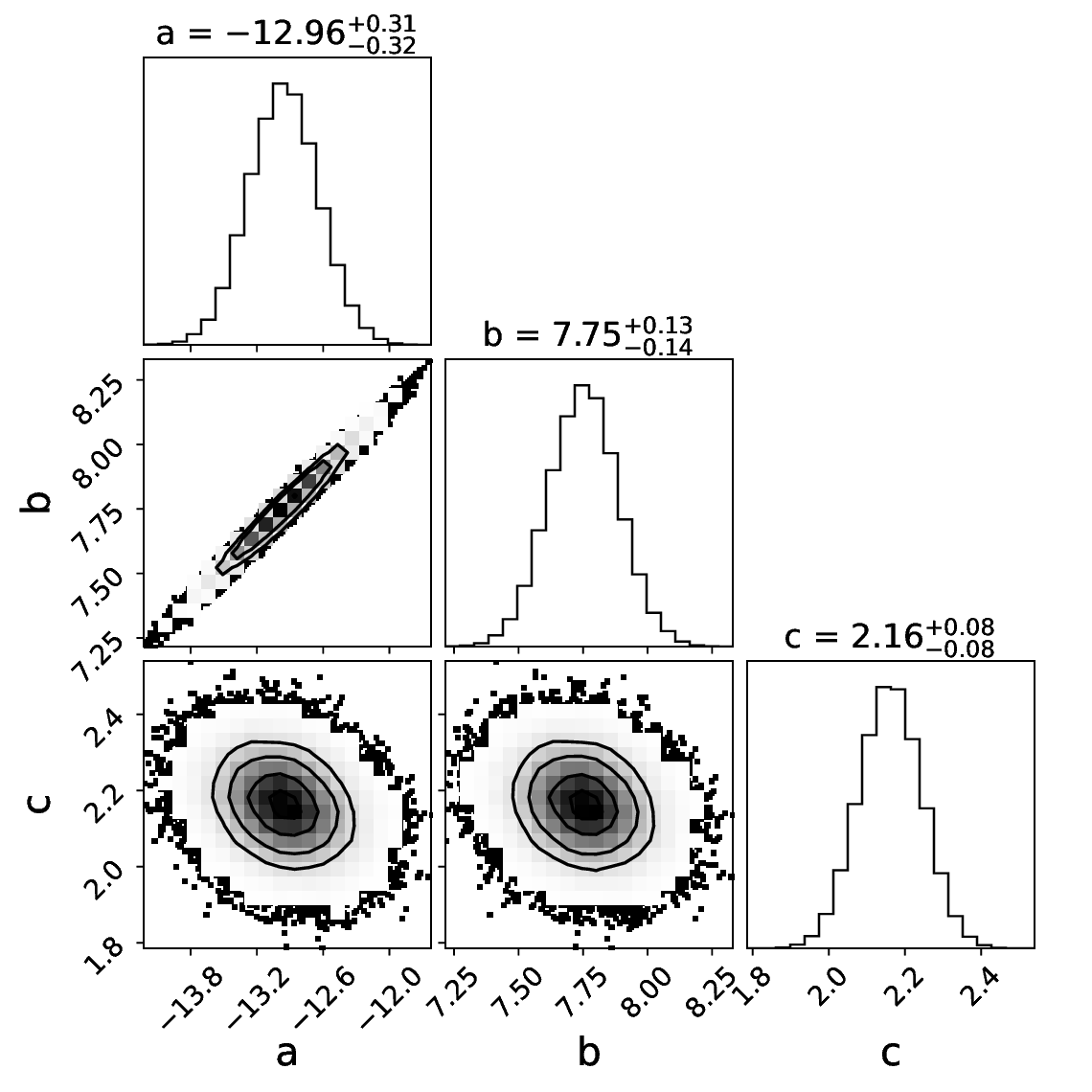}{0.33\textwidth}{(1) FUV-band}\hspace{-15mm}
         \fig{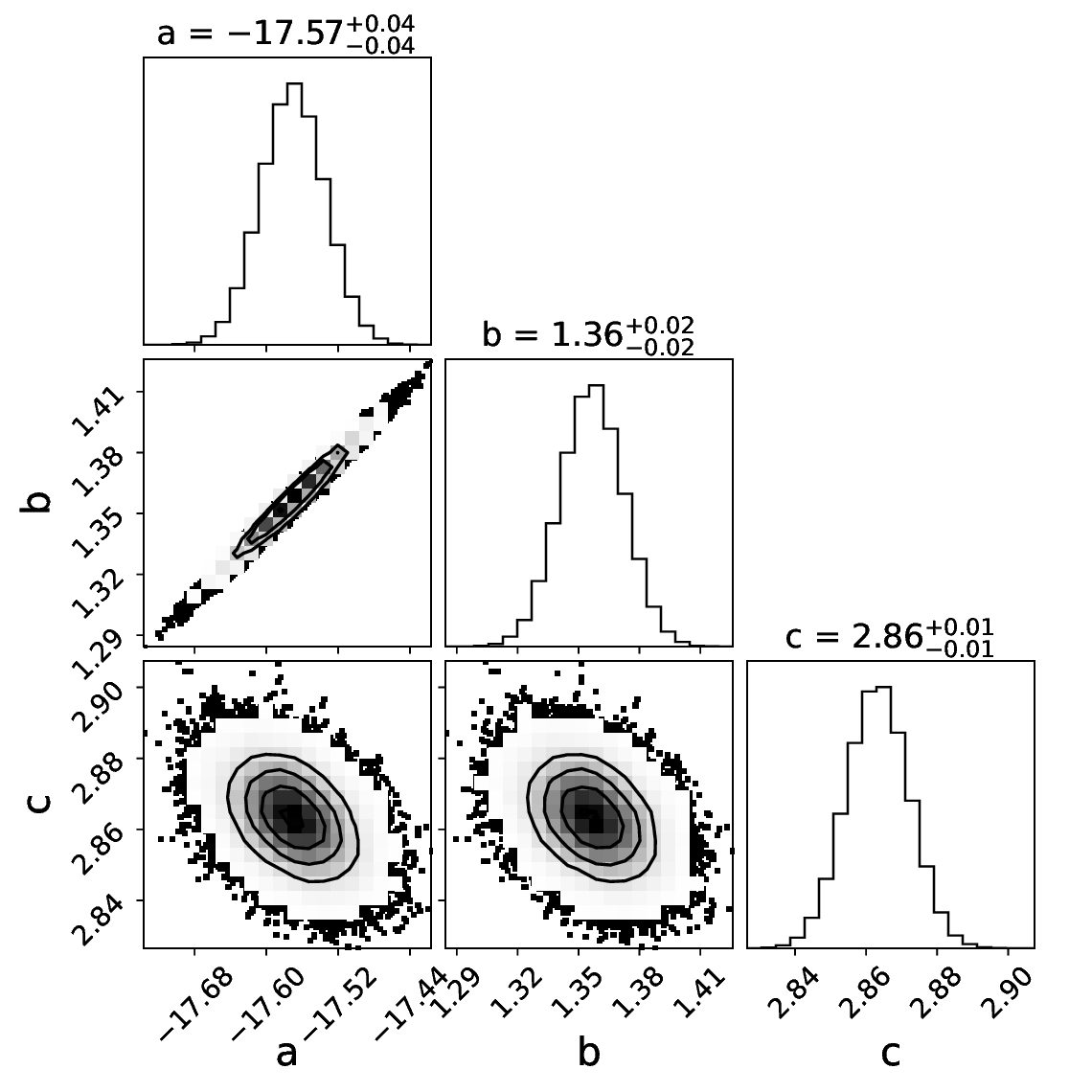}{0.33\textwidth}{(2) NUV-band}\hspace{-15mm}
          \fig{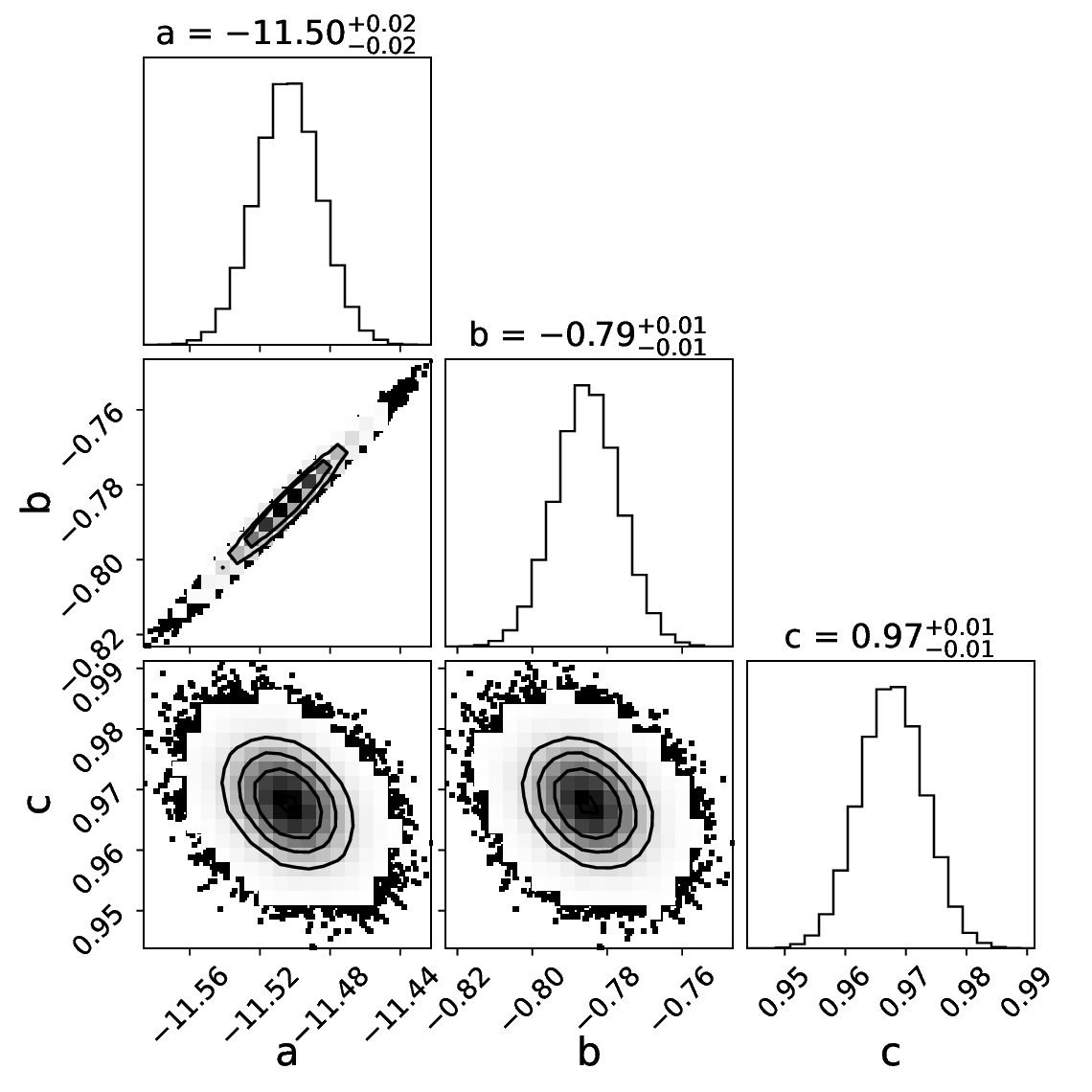}{0.33\textwidth}{(3) $B_{p}$-band}} 
 \gridline{\fig{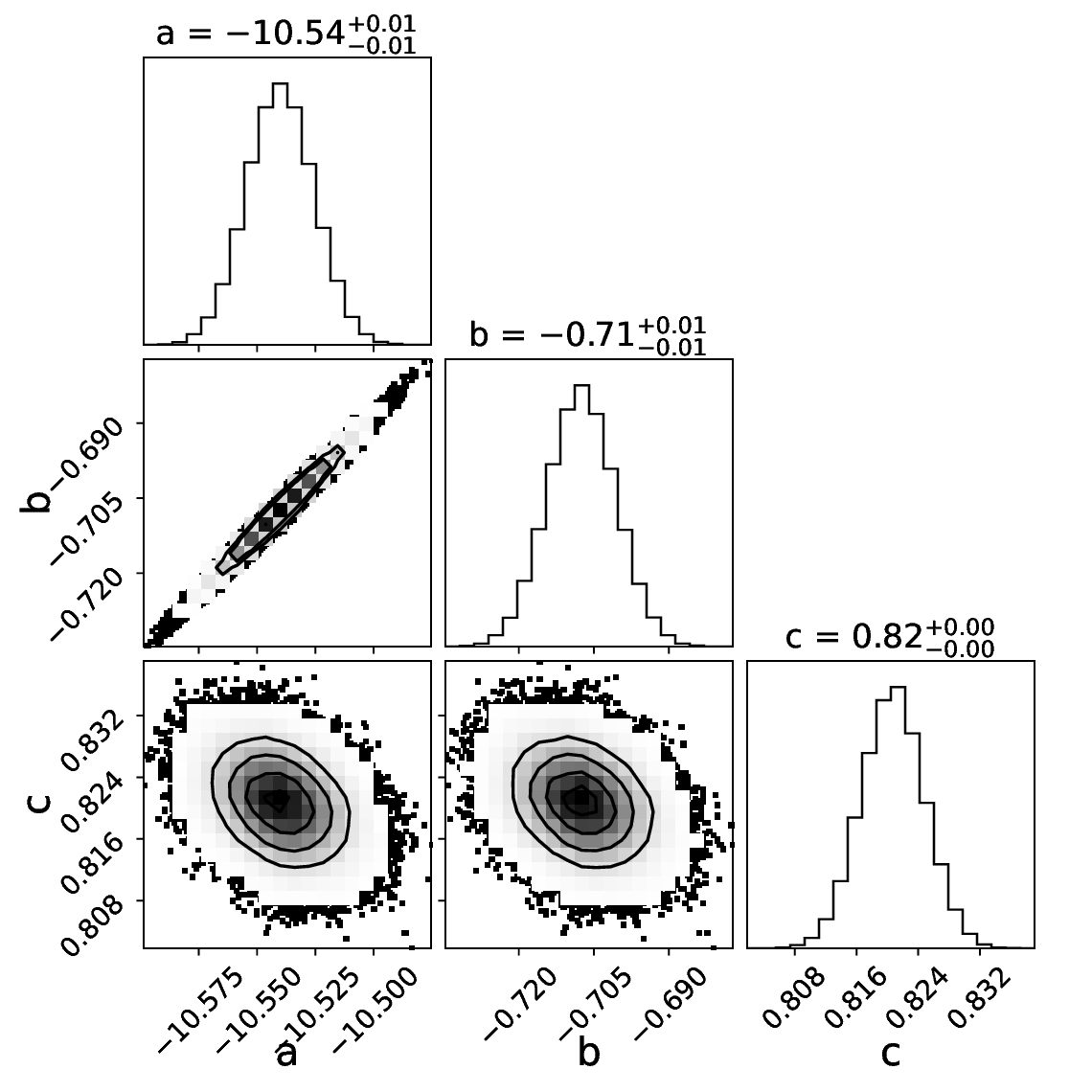}{0.33\textwidth}{(4) G-band}\hspace{-15mm}
           \fig{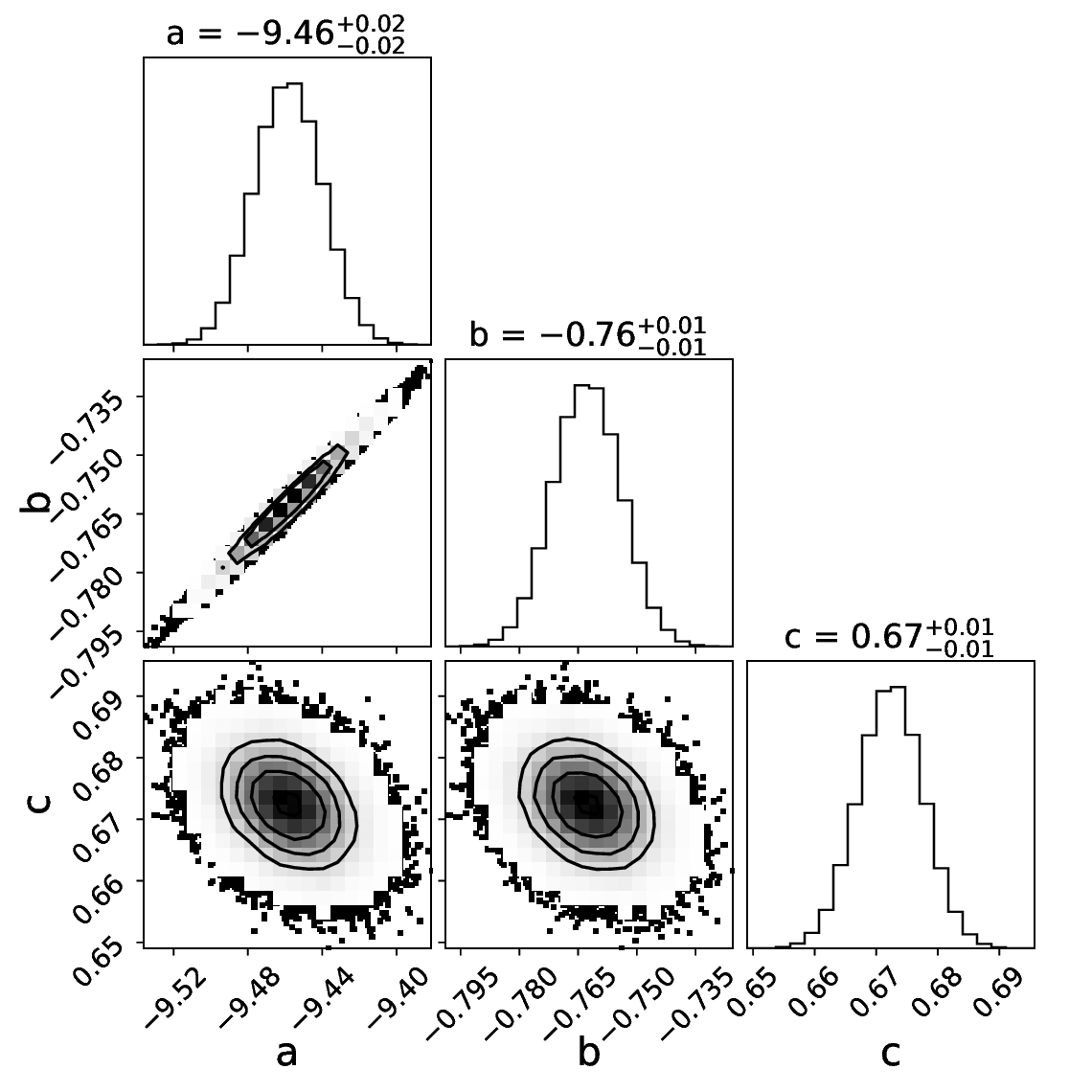}{0.33\textwidth}{(5) $R_{p}$-band}\hspace{-15mm}
          \fig{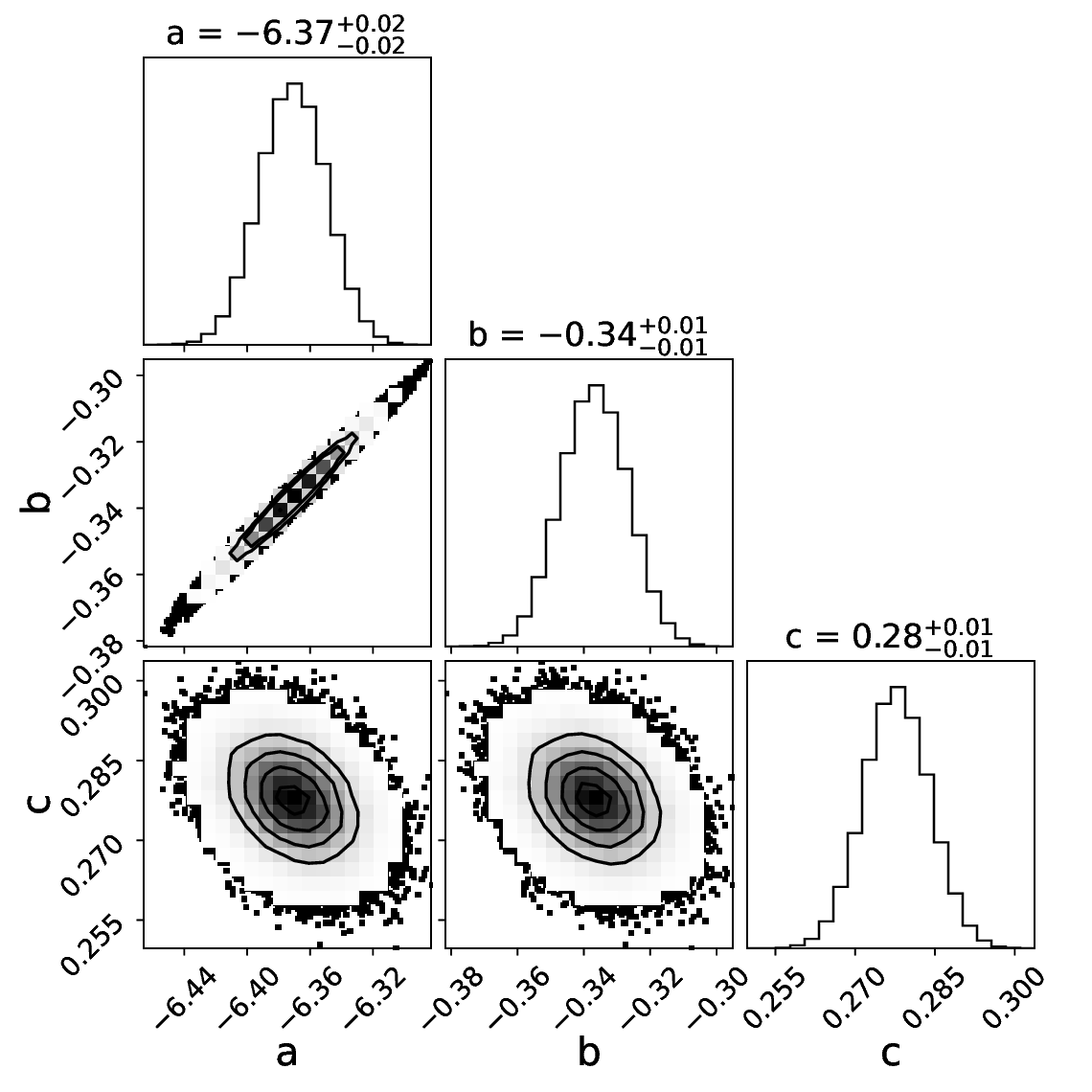}{0.33\textwidth}{(6) W1-band}}         
\gridline{\fig{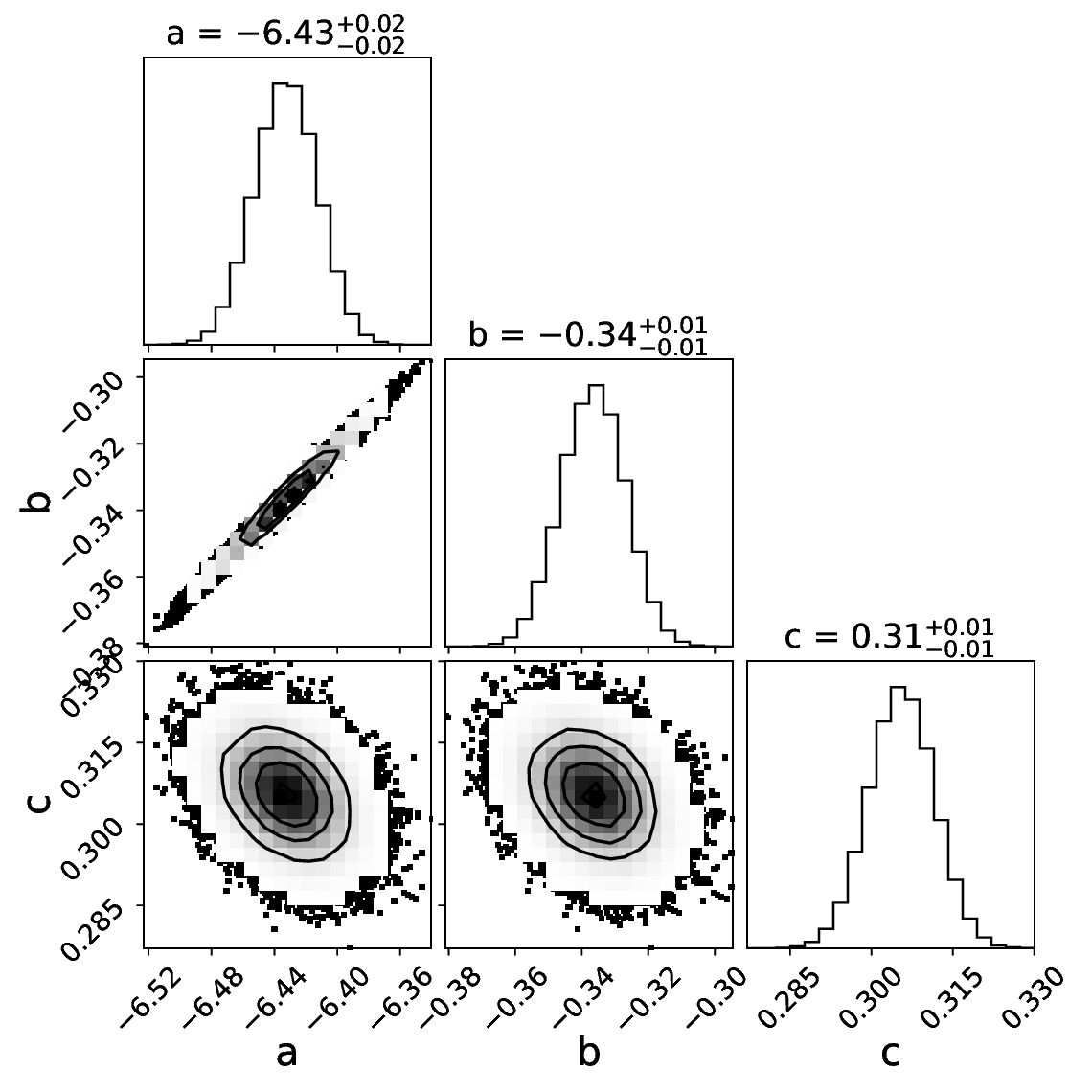}{0.33\textwidth}{(7) W2-band}\hspace{-15mm}
          \fig{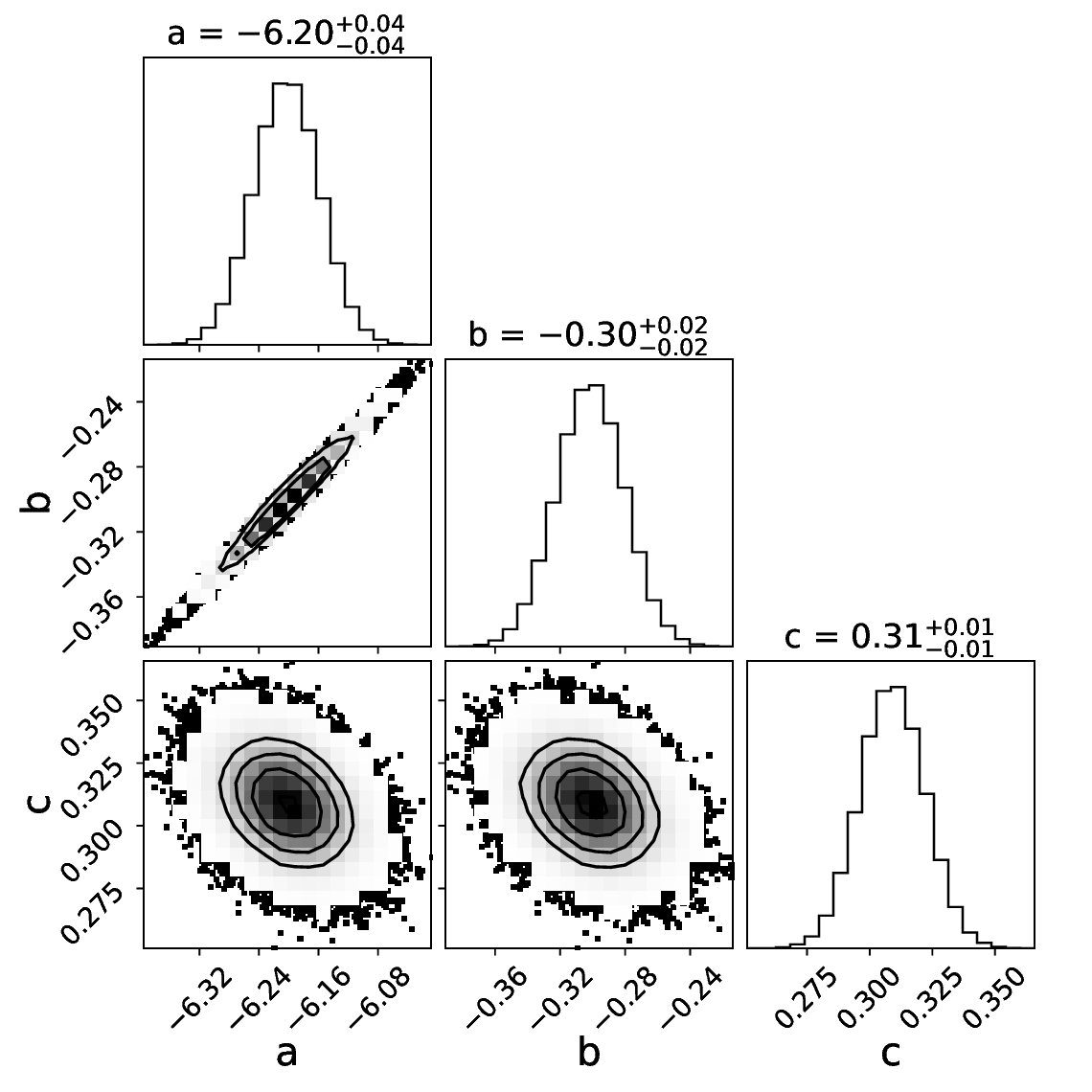}{0.33\textwidth}{(8) W3-band}\hspace{-15mm}
          \fig{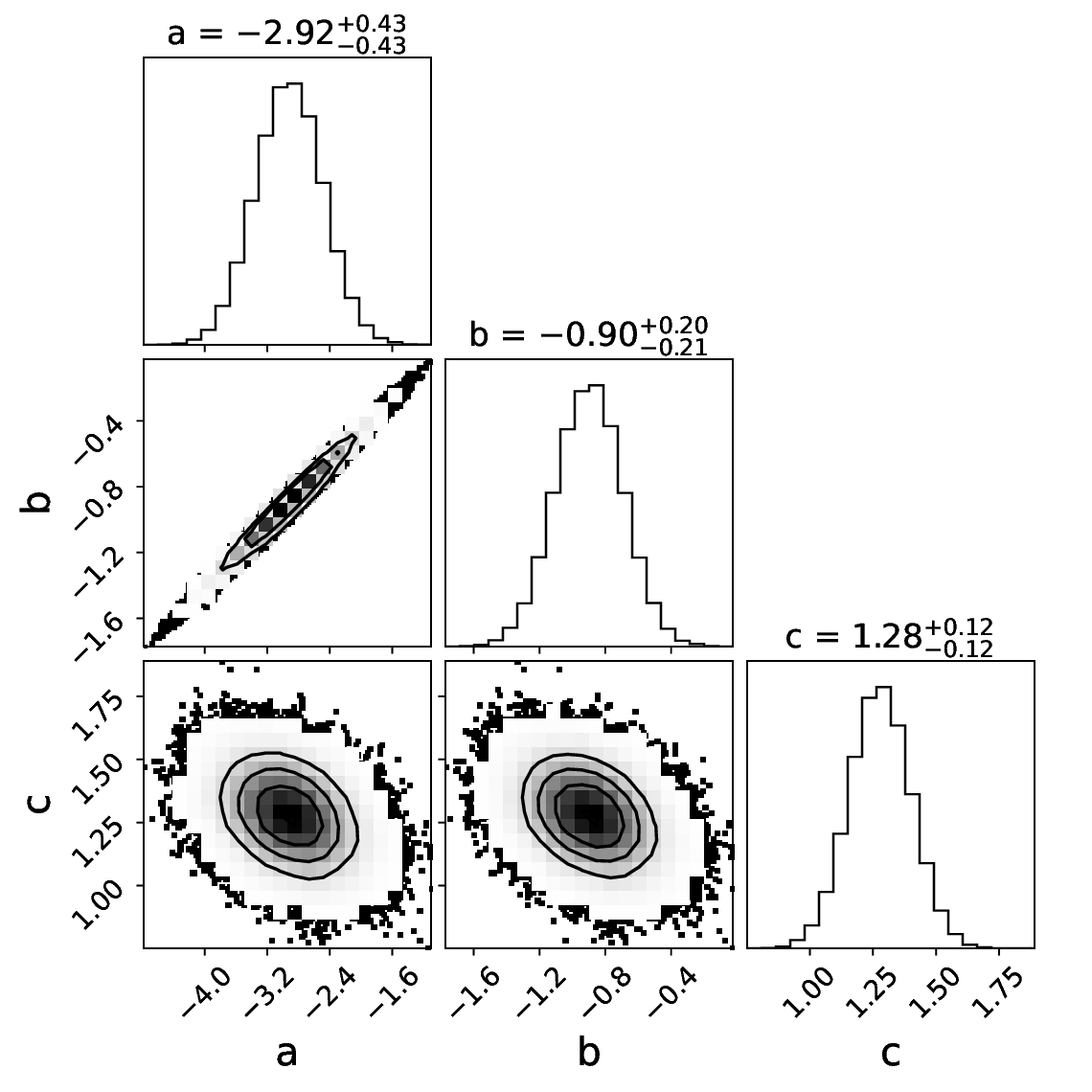}{0.33\textwidth}{(9) W4-band}}
\caption{Parameters fitting for PLZ relations by {\tt emcee}.
\label{fig.5}}
\end{figure*}

\begin{figure}
\gridline{\fig{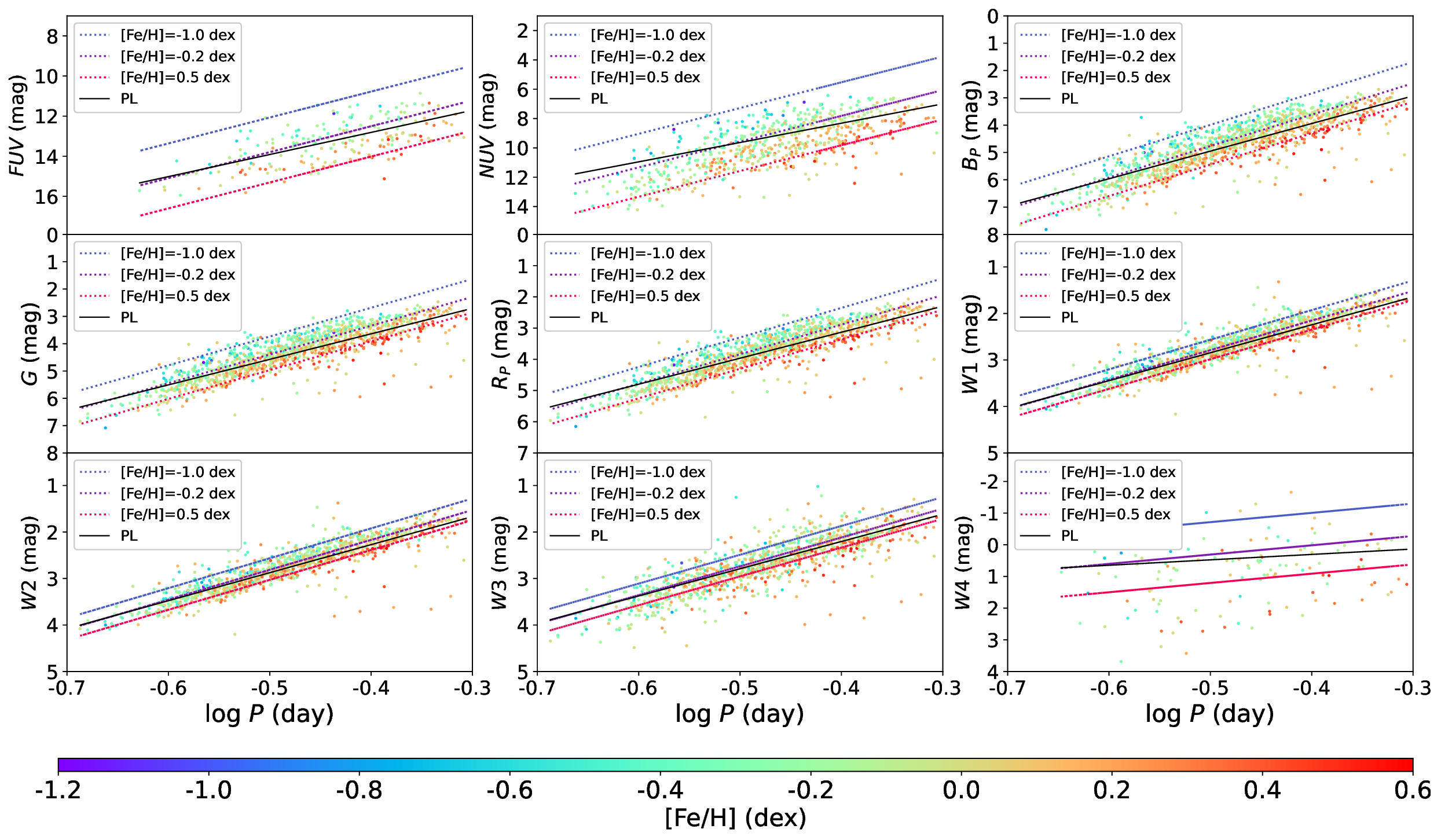}{1\textwidth}{}}  
\caption{The $\rm Period$-$\rm Magnitude$ diagram. The colorbar presents the value of $[\rm Fe/H]$.  
The dotted-lines in each panel present examples of PLZ relation with given $\rm [Fe/H]$, while the solid lines present the best-fitting PL relations.
\label{fig.6}}
\end{figure}

To verify the influence of metallicitiy abundance on PL relations, PLZ relations from ultraviolet to infrared bands are aslo derived.
Corner plots of {\tt emcee} fitting of equation \ref{equ2} are presented in Fig. \ref{fig.5}.  
Table \ref{table3} summarizes their fitting parameters and responding dispersion. 
The dispersions are improved when considering the influence of [Fe/H] on PL relations. 
With 1.0 dex increasing of $\rm [Fe/H]$, the CBs will be dimmer of 0.28 - 2.86 mag 
for the ultraviolet to infrared bands.
It is slightly higher than 0.2 mag proposed by \citet{CDD+2016}. 
To present the fitting results of the PLZ relations on $\rm Period$-$\rm Magnitude$ diagram,  
we choose three values, -1.0, -0.2, and 0.5 dex of $\rm [Fe/H]$, as examples in Fig. \ref{fig.6}.
It sees from the figure that the PLZ relations fit the observations better than hte PL relations.
\begin{deluxetable}{cccccch}
\setlength{\tabcolsep}{7mm}
\tablecaption{Parameters of PLZ models for multi-band. \label{table3}}
\tablewidth{5pt}
\tablehead{
Band & Number &  a & b & c & $\sigma$ (mag) }
\startdata
$FUV$     & 199    & $-$12.96($\pm0.32$) & 7.75($\pm0.13$)  & 2.16($\pm0.08$)  & 0.71 \\
$NUV$     & 632    & $-$17.57($\pm0.02$) & 1.36($\pm0.02$)  & 2.86($\pm0.01$)  & 0.82 \\
\hline
$B_{p}$  & 1,093  & $-$11.50($\pm0.02$) & $-$0.79($\pm0.01$) & 0.97($\pm0.01$) & 0.28 \\
$G$      & 1,093  & $-$10.54($\pm0.02$) & $-$0.71($\pm0.01$) & 0.82($\pm0.01$) & 0.25\\
$R_{p}$  & 1,093  & $-$9.46($\pm0.02$)  & $-$0.76($\pm0.01$) & 0.67($\pm0.01$) & 0.22 \\
\hline
$W1$    & 1,019  & $-$6.37($\pm0.02$)  & $-$0.34($\pm0.01$)  & 0.28($\pm0.01$) & 0.12\\
$W2$    & 1,019  & $-$6.43($\pm0.02$)  & $-$0.34($\pm0.01$)  & 0.31($\pm0.01$) & 0.14 \\
$W3$    & 992    & $-$6.20($\pm0.04$)  & $-$0.30($\pm0.02$)  & 0.31($\pm0.01$) & 0.22\\
$W4$    & 127    & $-$2.92($\pm0.20$)  & $-$0.90($\pm0.09$)  & 1.28($\pm0.07$) & 1.40  \\
\enddata
\tablecomments{The parameters a, b, and c are best-fitting coefficients in equation \ref{equ2}, 
and $\sigma$ denote the deviations of PLZ relations.}
\end{deluxetable}

\begin{figure*}
\gridline{\fig{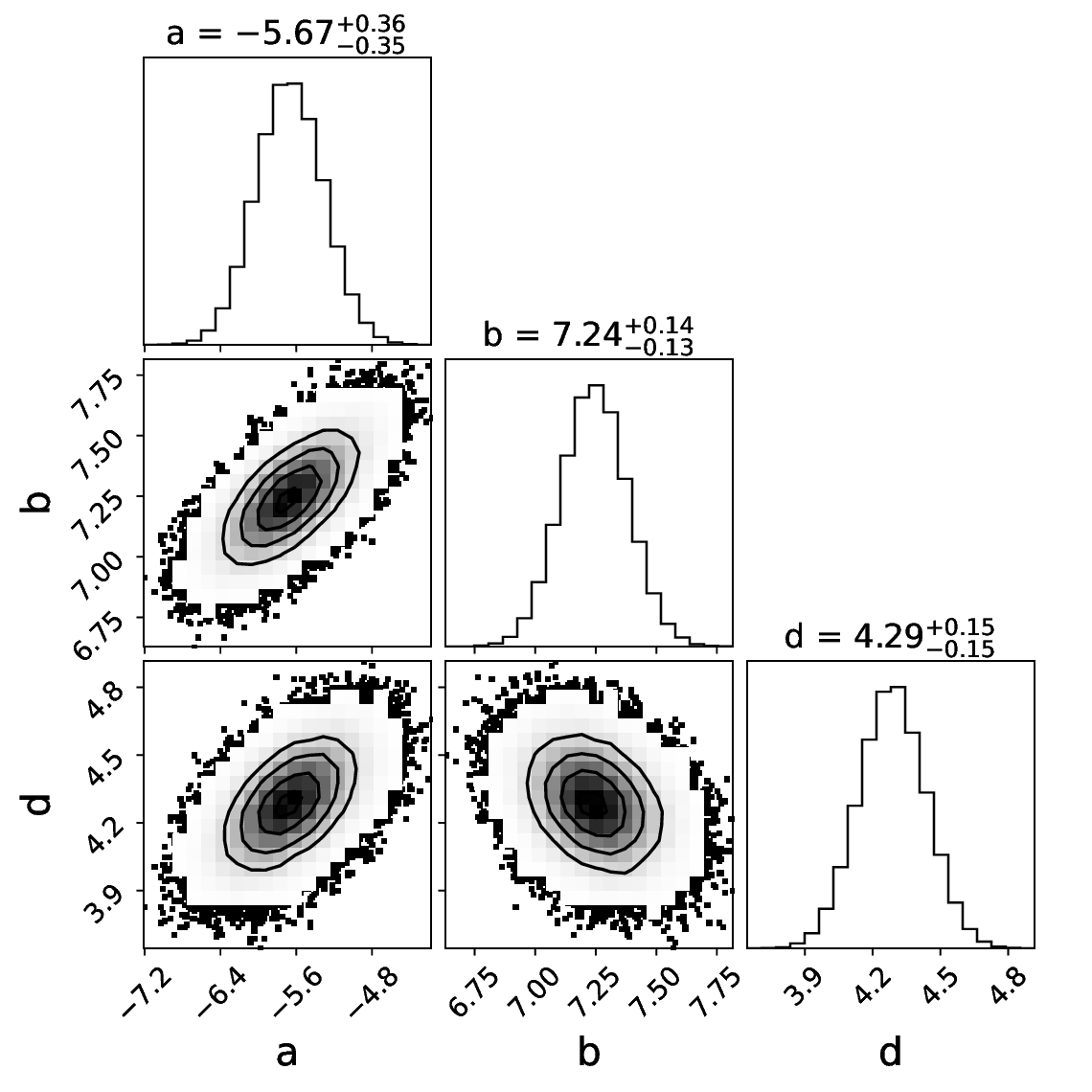}{0.32\textwidth}{(1) FUV-band}\hspace{-15mm} 
         \fig{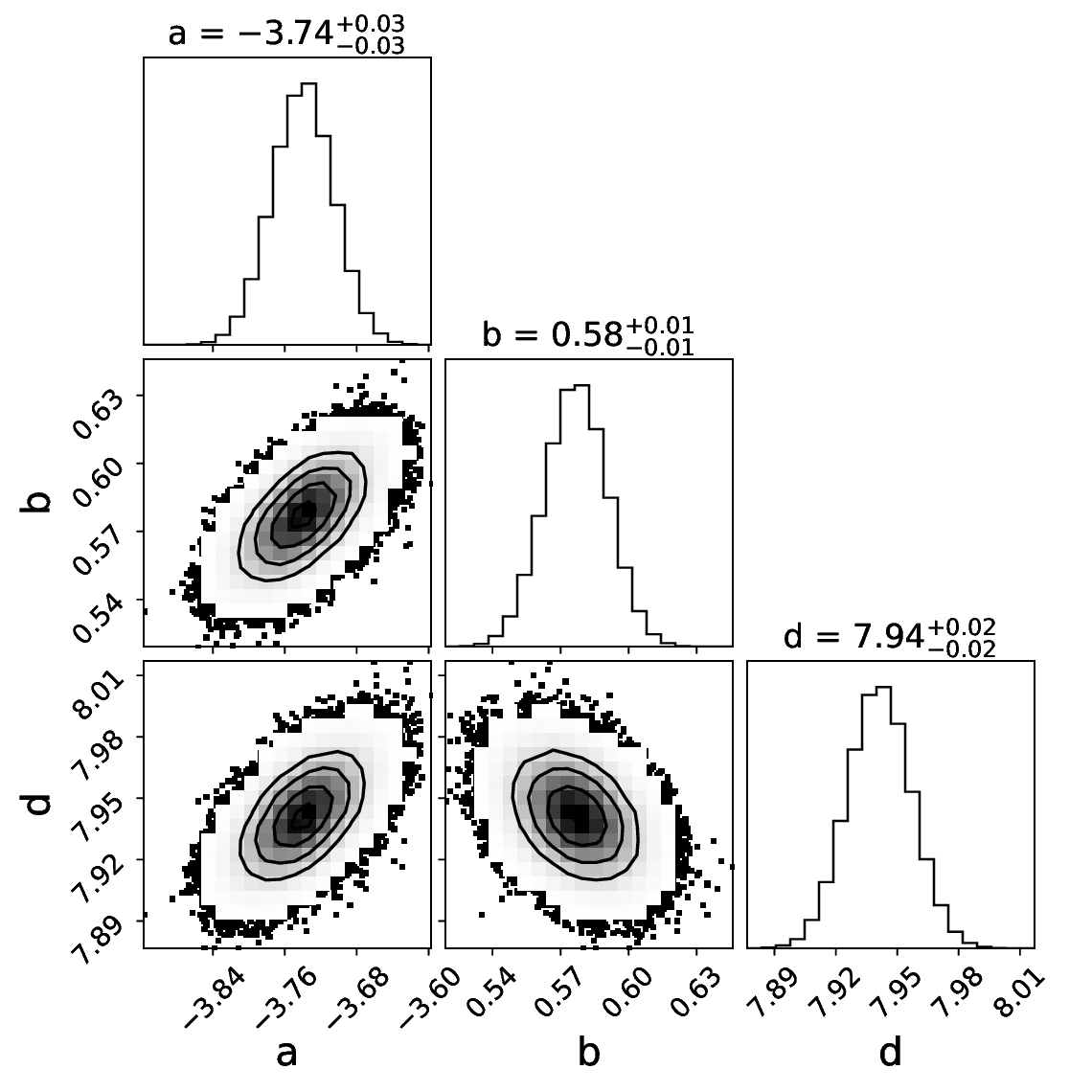}{0.32\textwidth}{(2) NUV-band}\hspace{-15mm}
          \fig{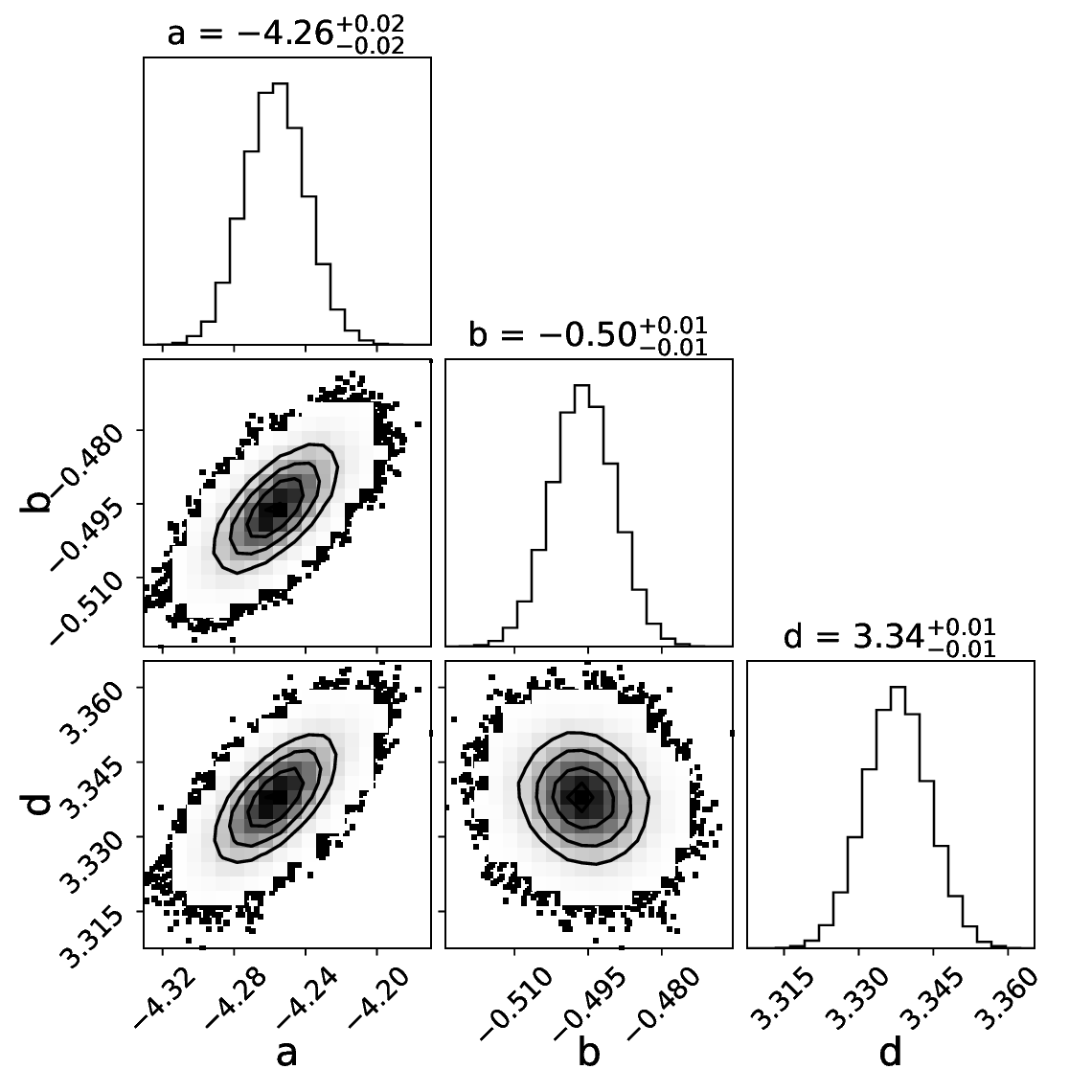}{0.32\textwidth}{(3) $B_{p}$-band}}
 \gridline{\fig{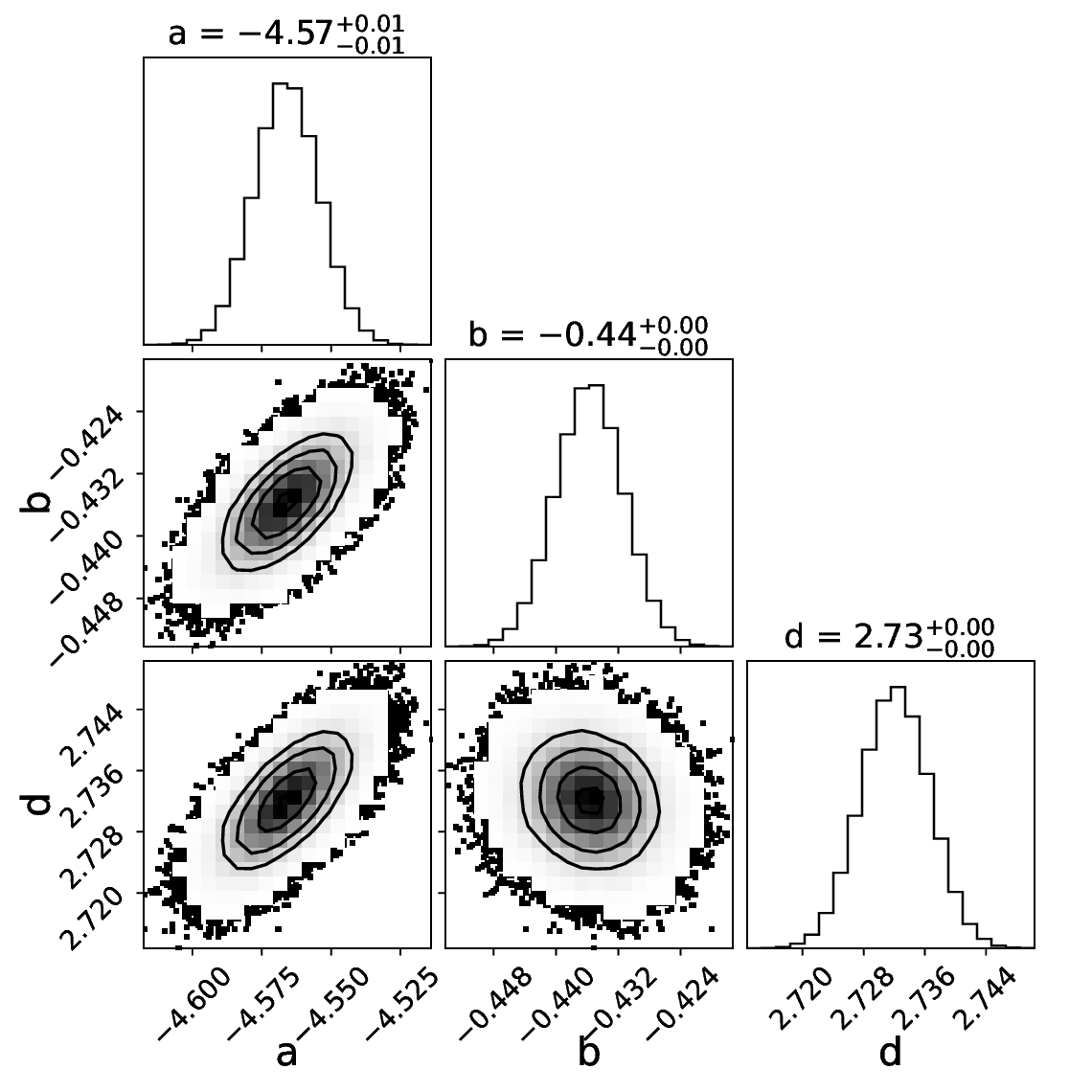}{0.32\textwidth}{(4) G-band}\hspace{-15mm}
           \fig{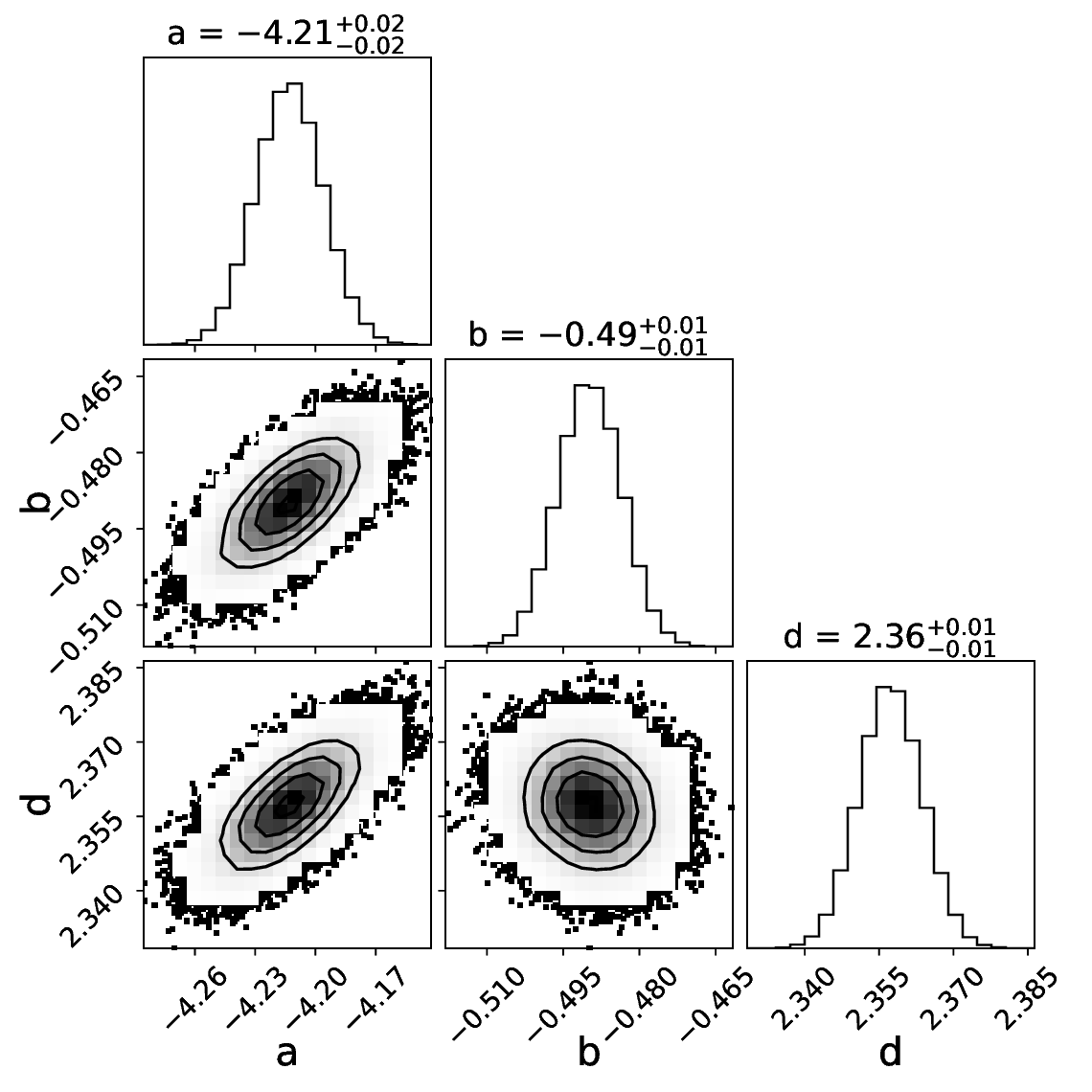}{0.32\textwidth}{(5) $R_{p}$-band}\hspace{-15mm}
          \fig{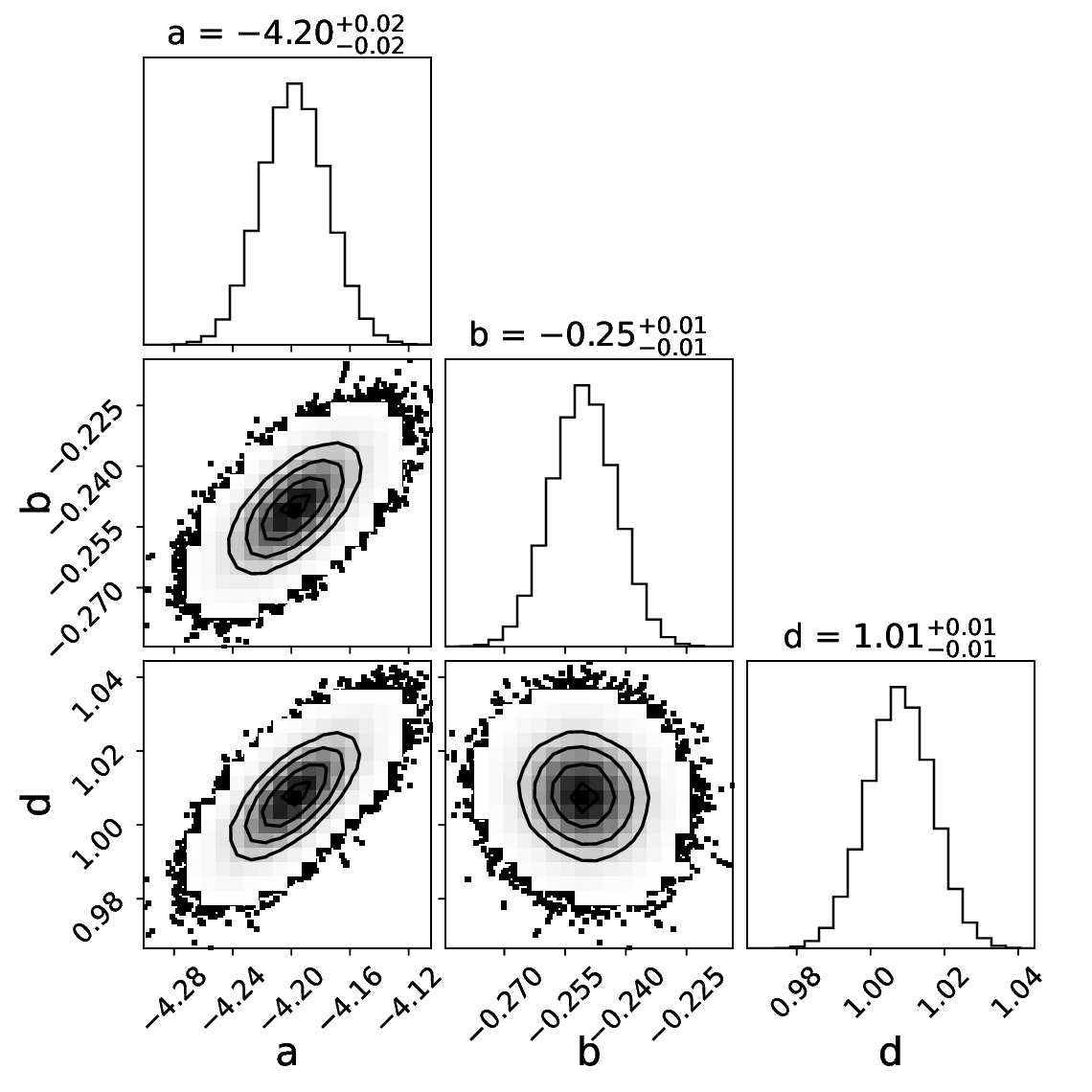}{0.32\textwidth}{(6) W1-band}}         
\gridline{\fig{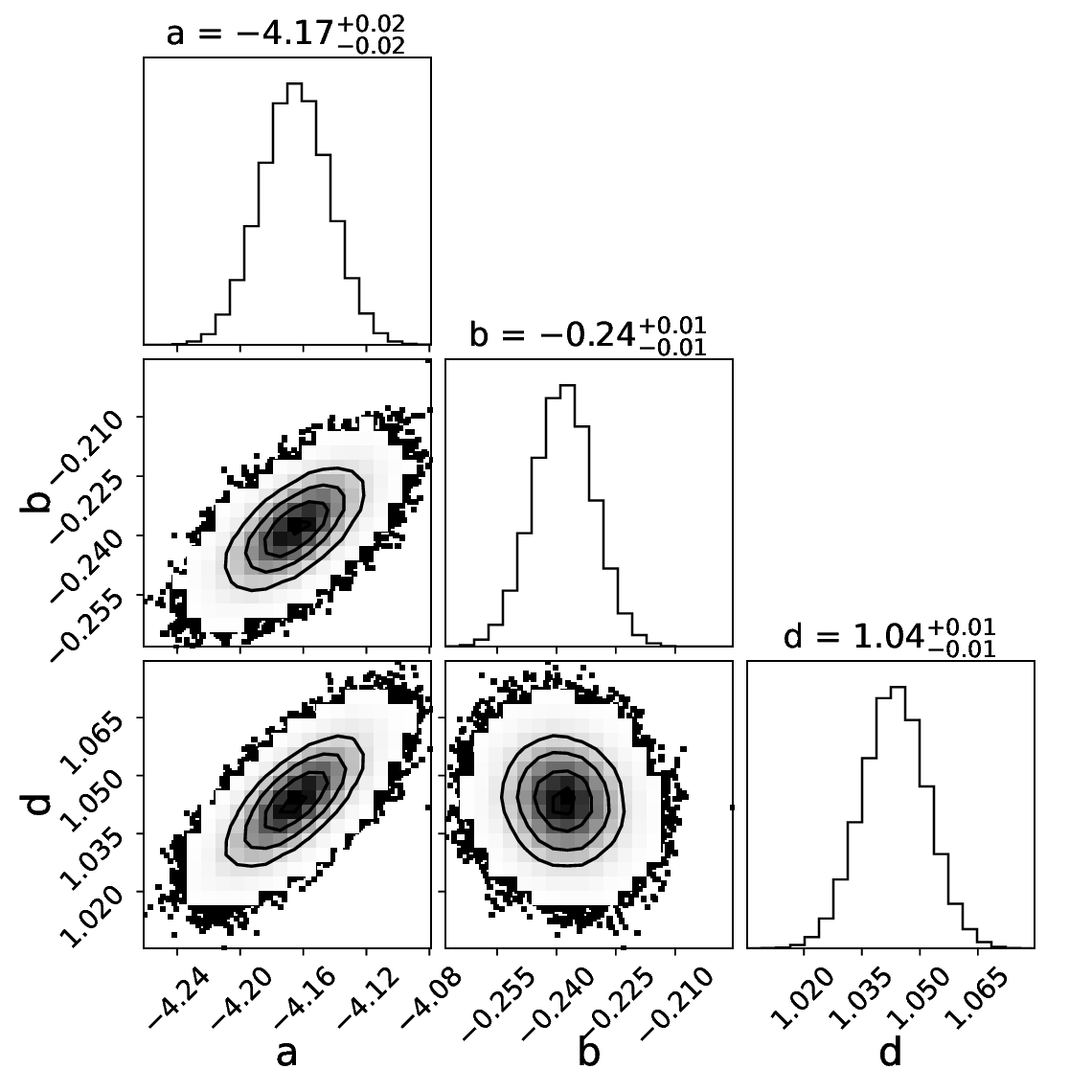}{0.32\textwidth}{(7) W2-band}\hspace{-15mm}
          \fig{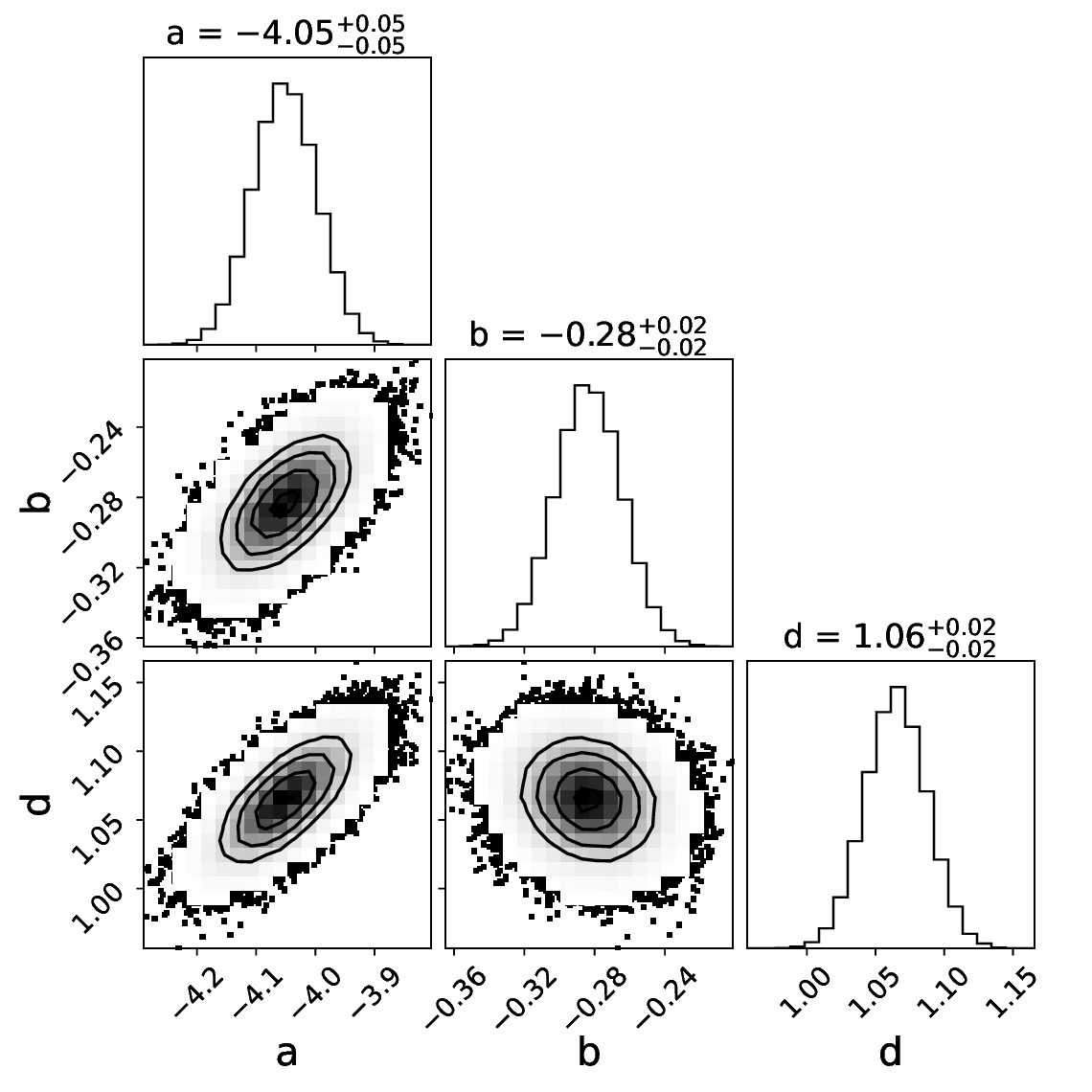}{0.32\textwidth}{(8) W3-band}\hspace{-15mm}
          \fig{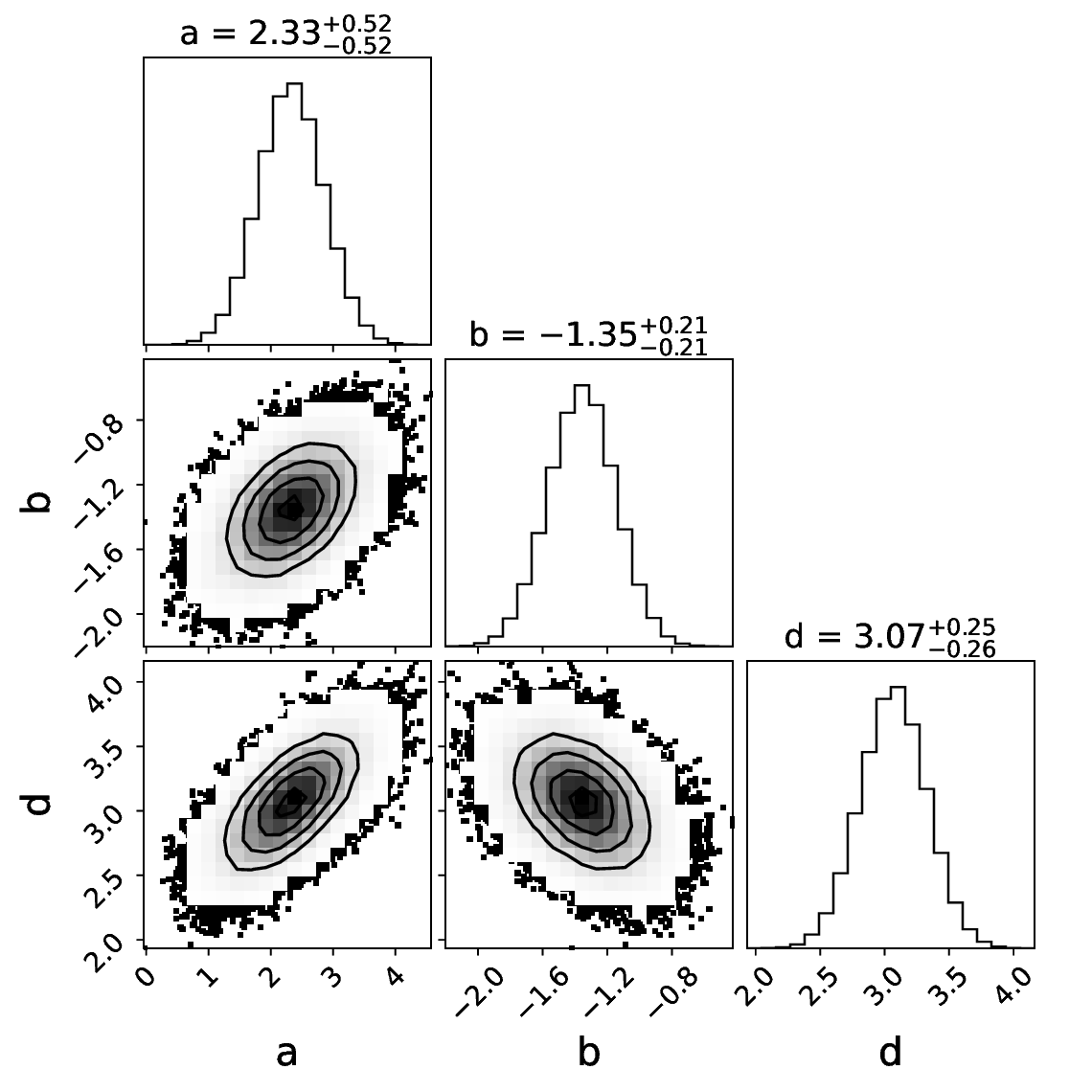}{0.32\textwidth}{(9) W4-band}}
\caption{Parameters fitting for PLC relations by {\tt emcee}.
\label{fig.7}}
\end{figure*}

\begin{figure}
\gridline{\fig{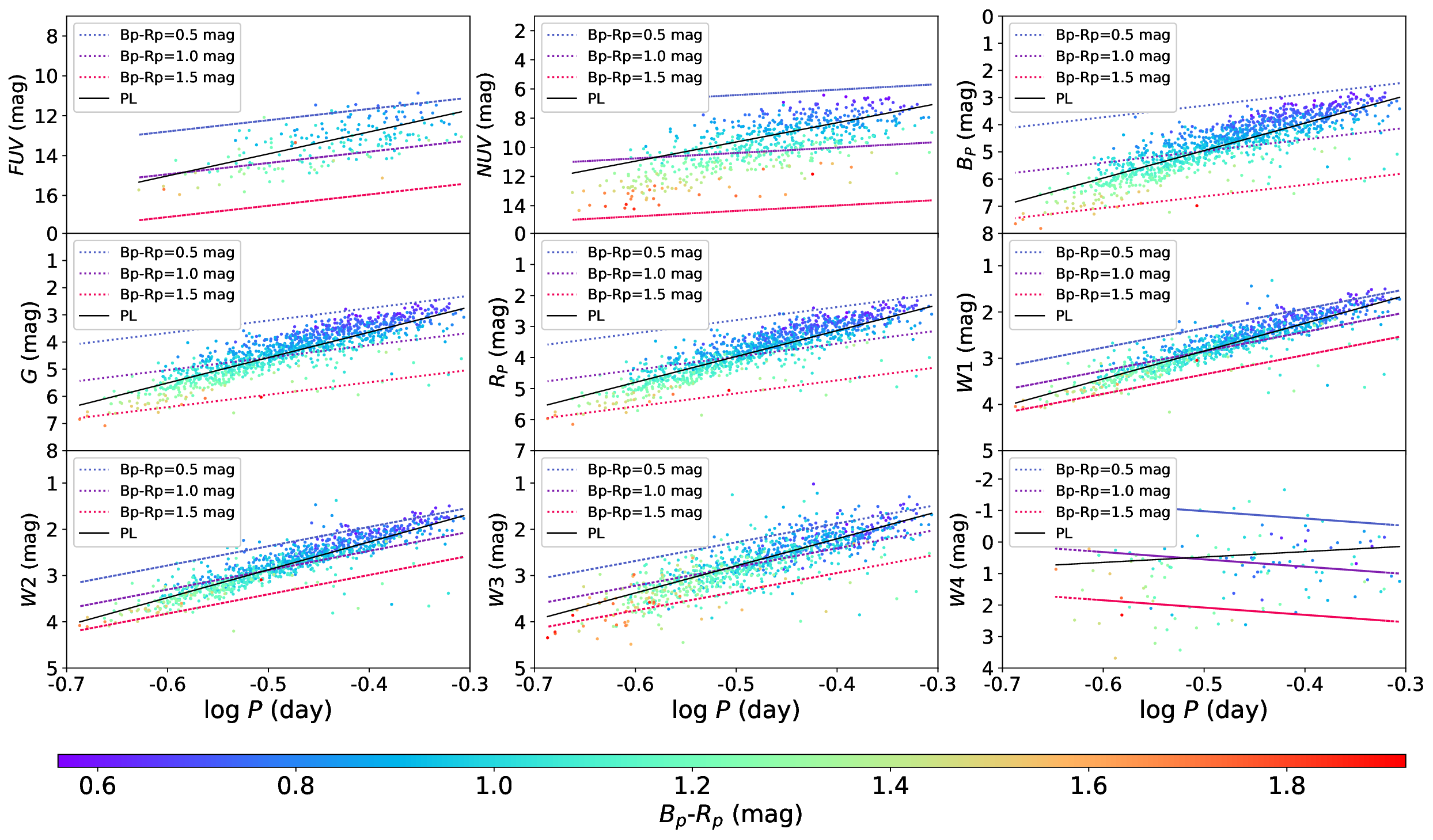}{1\textwidth}{}} 
\caption{The $\rm Period$-$\rm Magnitude$ diagram. The colorbar denotes the $B_{p}-R_{p}$ color. The dotted-lines in each panel present examples of PLC relation with given colors, while the solid lines present the best-fitting PL relations.
\label{fig.8}}
\end{figure}

According to Roche-lobe theory \citep{1983ApJ...268..368E},
it is anticipated that the correlation between the CBs orbital period and absolute magnitudes will include both the effective temperature of the system and the color term.
The PLC relations for multi bands by using equation \ref{equ3} are calibrated, of which the performance are shown in Fig. \ref{fig.7}.
Table \ref{table4} summarizes their fitting parameters and dispersion. 
Similarly, we also show the result of PLC relations on Fig. \ref{fig.8}. 
The slope of $\log P$ in the PLC relations are lower in comparison to the PL and PLZ relations, which proves the contribution of color term to calibrations. 
The figure shows that the magnitudes of CBs are more sensitive to color term than metallicitiy term.
\begin{deluxetable}{cccccch}
\setlength{\tabcolsep}{7mm}
\tablecaption{Parameters of PLC models for multi-band. \label{table4}}
\tablewidth{5pt}
\tablehead{
Band & Number &  a & b & d & $\sigma$ (mag) }
\startdata
$FUV$ & 199  & $-$5.67($\pm0.36$) & 7.24($\pm0.13$)  & 4.29($\pm0.15$)  & 0.61 \\
$NUV$ & 632  & $-$3.74($\pm0.03$)  & 0.58($\pm0.01$) & 7.94($\pm0.02$) & 0.36 \\
\hline
$B_{p}$  & 1,093  & $-$4.26($\pm0.02$)  & $-$0.50($\pm0.01$) & 3.34($\pm0.01$)& 0.19 \\
$G$  & 1,093  & $-$4.57($\pm0.01$) & $-$0.44($\pm0.01$)  & 2.73($\pm0.01$) & 0.18\\
$R_{p}$  & 1,093  & $-$4.21$\pm0.02$)  & $-$0.49($\pm0.01$) & 2.36($\pm0.01$)& 0.19 \\
\hline
$W1$  & 1,019 & $-$4.20($\pm0.02$)  & $-$0.25($\pm0.01$) & 1.01($\pm0.01$)& 0.12\\
$W2$  & 1,019  & $-$4.17($\pm0.02$) & $-$0.24($\pm0.01$)  & 1.04($\pm0.01$) & 0.12 \\
$W3$  &992  & $-$4.05($\pm0.05$)  & $-$0.28($\pm0.02$) & 1.06($\pm0.02$)& 0.22\\
$W4$  & 127 & 2.33($\pm0.52$) & $-$1.35($\pm0.21$)  & 3.07($\pm0.25$) & 1.15  \\
\enddata
\tablecomments{The parameters a, b, and d are best-fitting coefficients in equation \ref{equ3}, 
and $\sigma$ denote the deviations of PLC relations.}
\end{deluxetable}

\begin{figure*}
\gridline{\fig{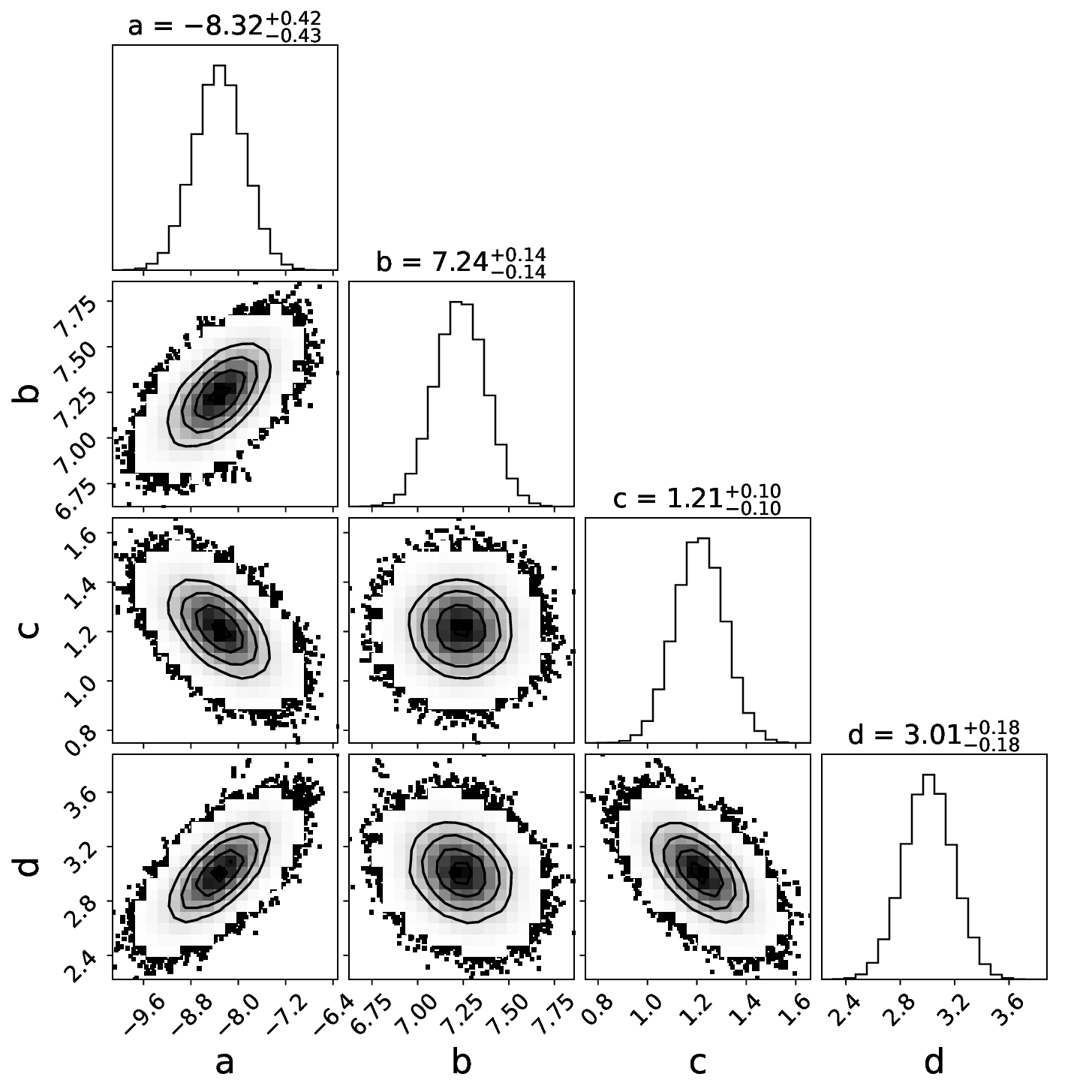}{0.325\textwidth}{(1) FUV-band}\hspace{-15mm} 
         \fig{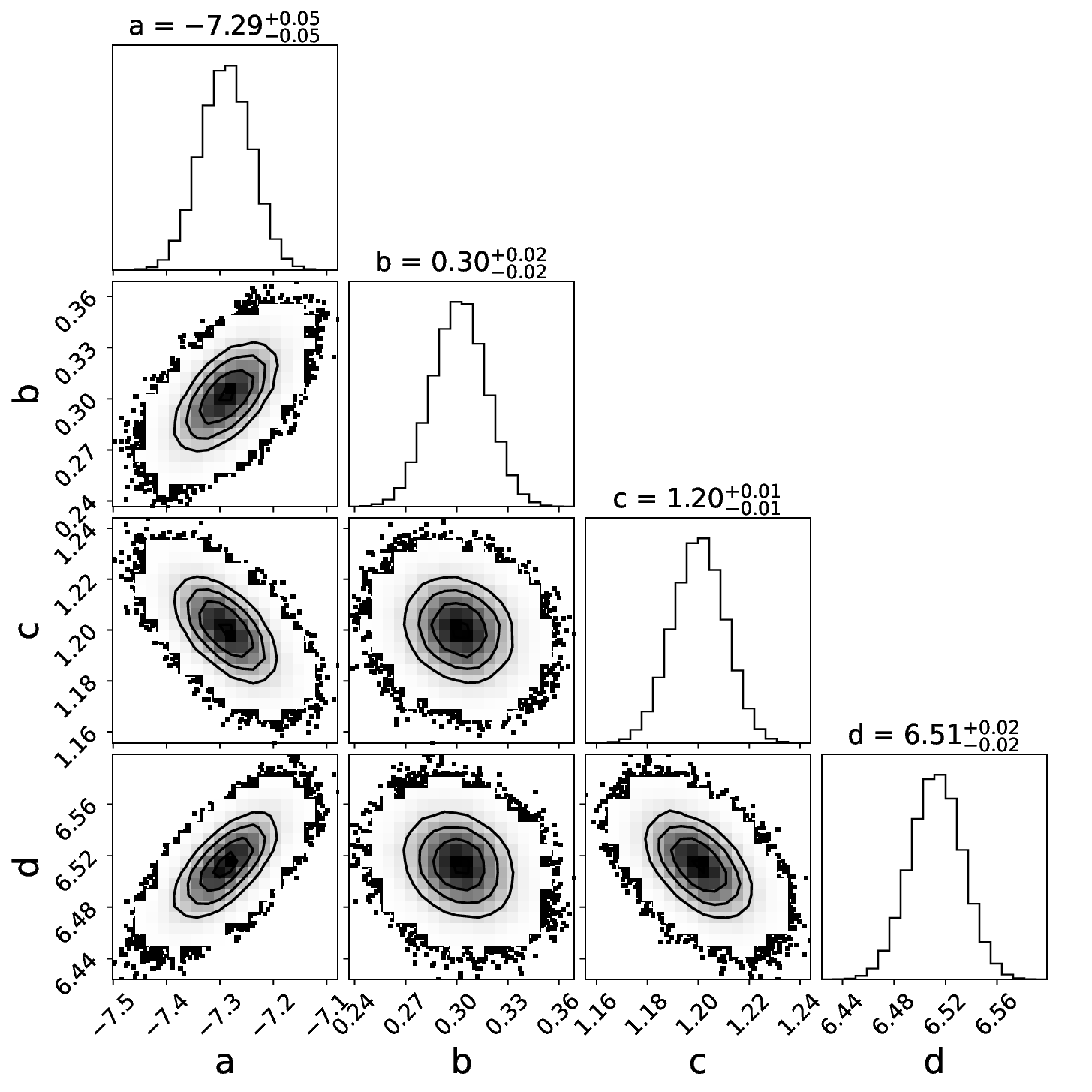}{0.325\textwidth}{(2) NUV-band}\hspace{-15mm} 
          \fig{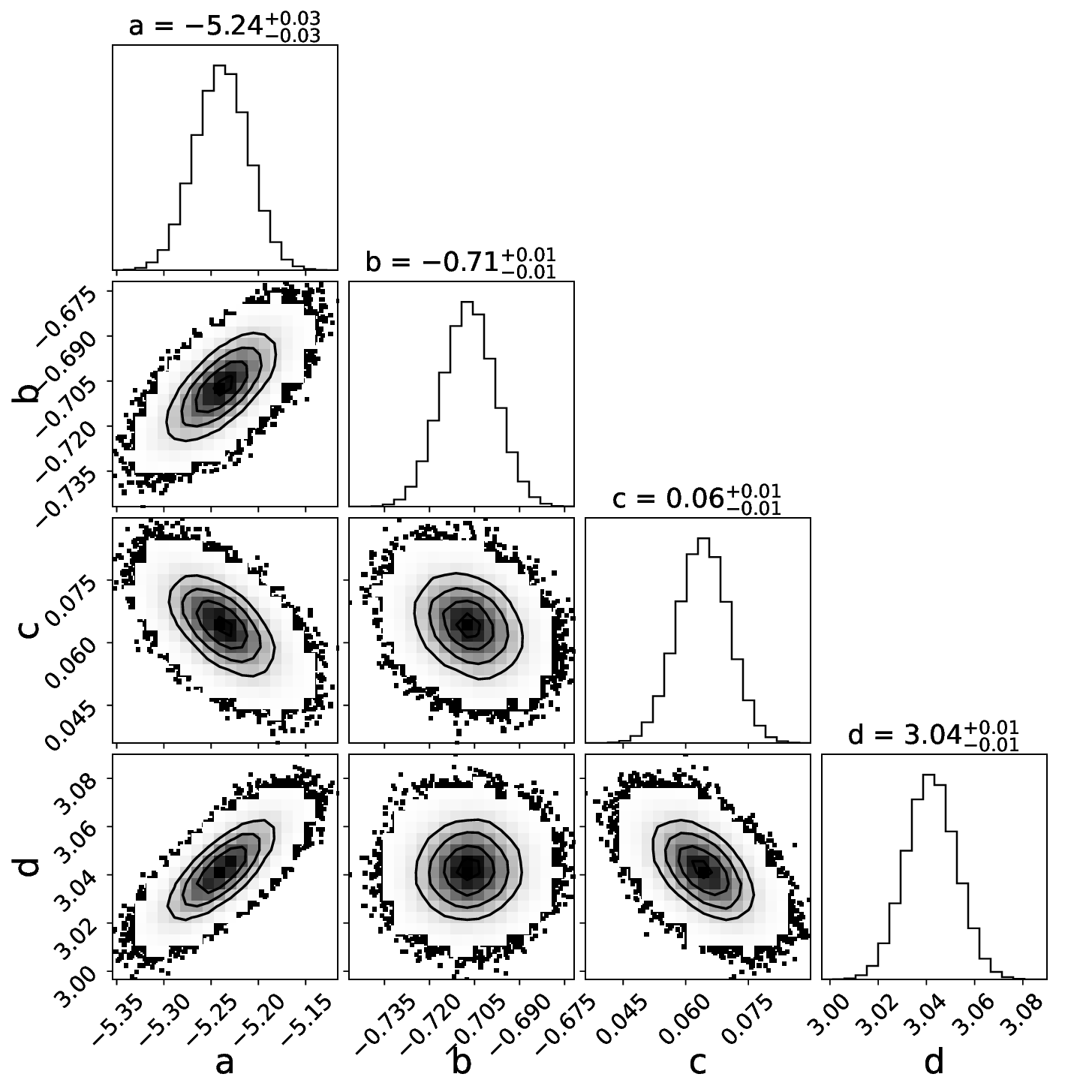}{0.325\textwidth}{(3) $B_{p}$-band}}
 \gridline{\fig{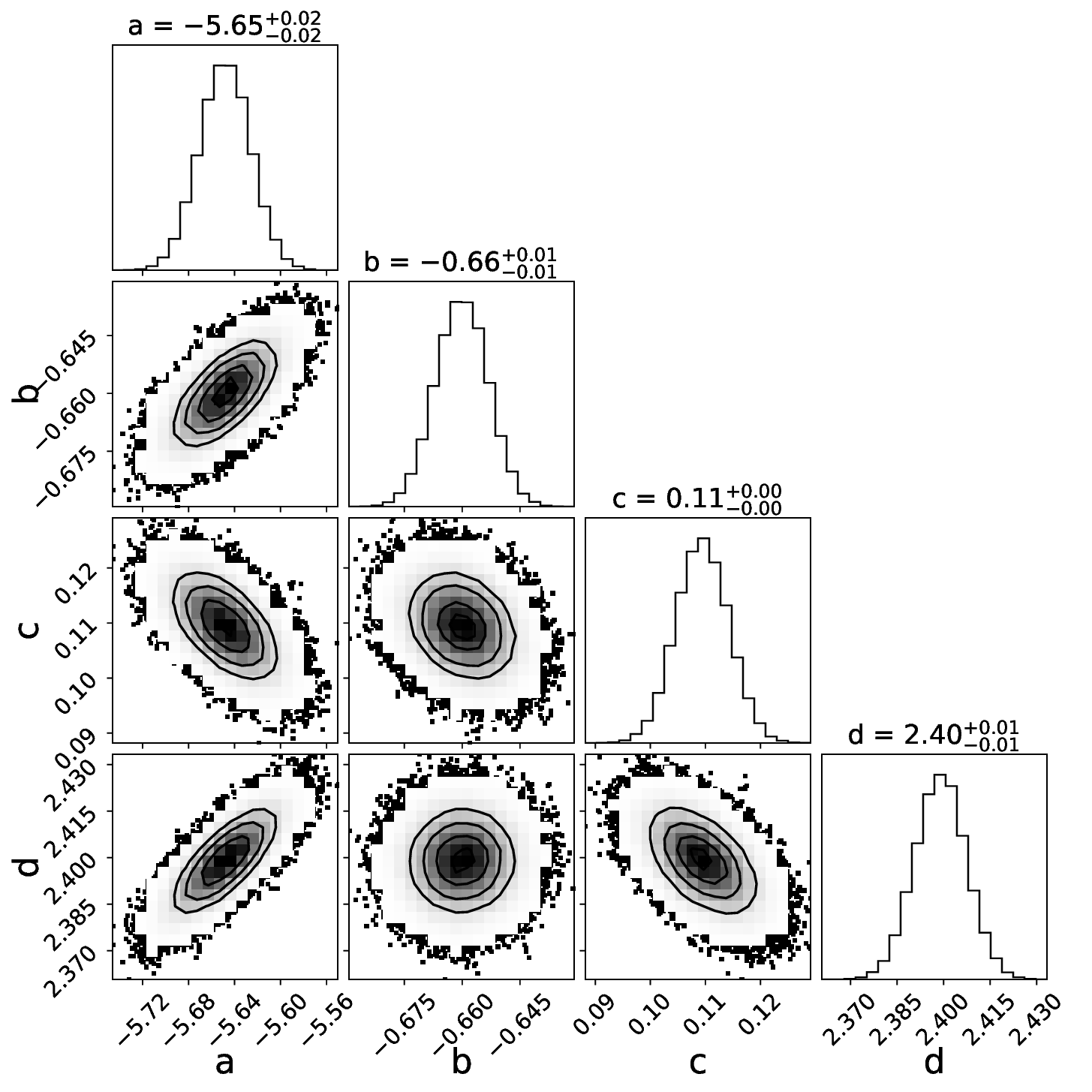}{0.325\textwidth}{(4) G-band}\hspace{-15mm} 
           \fig{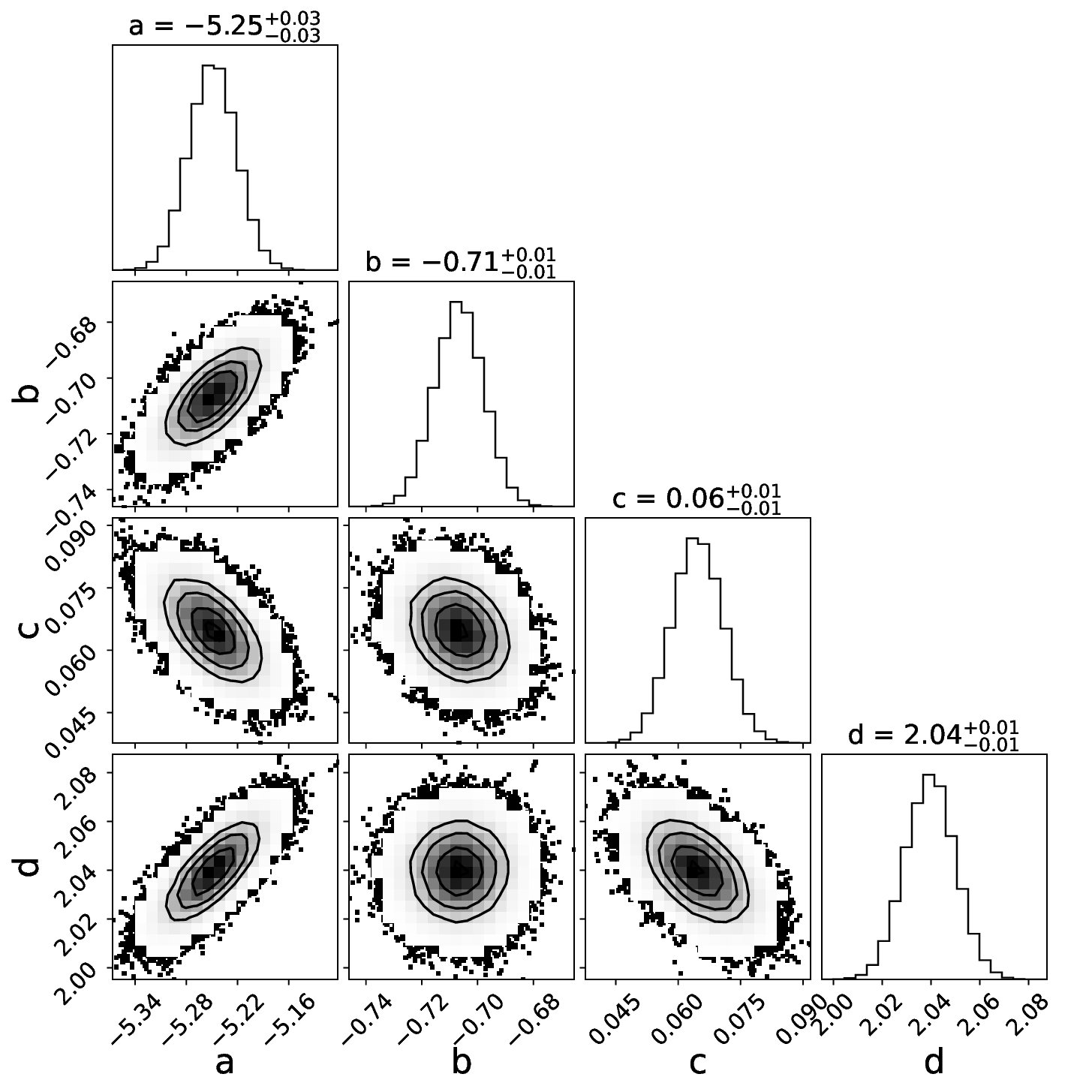}{0.325\textwidth}{(5) $R_{p}$-band}\hspace{-15mm} 
          \fig{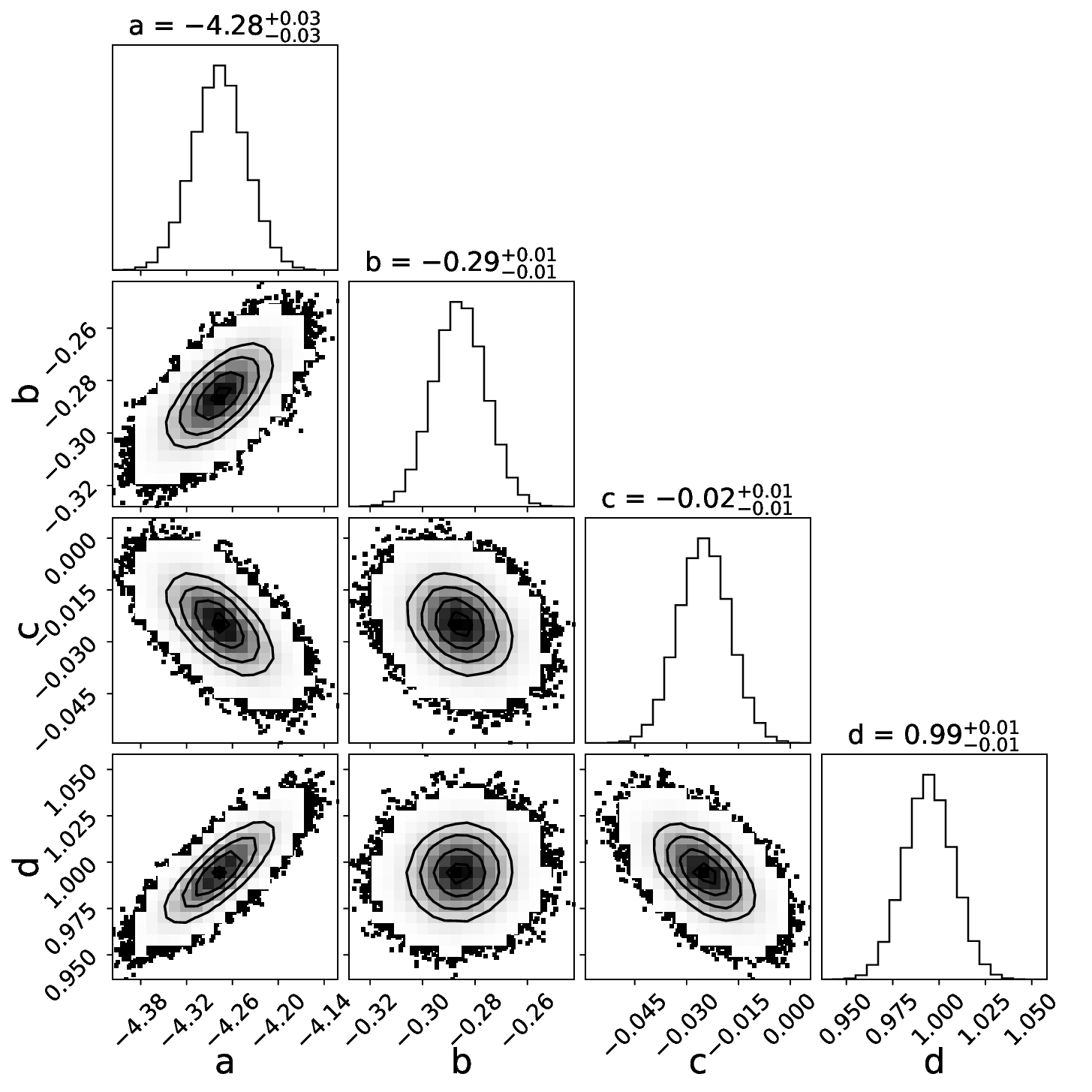}{0.325\textwidth}{(6) W1-band}}         
\gridline{\fig{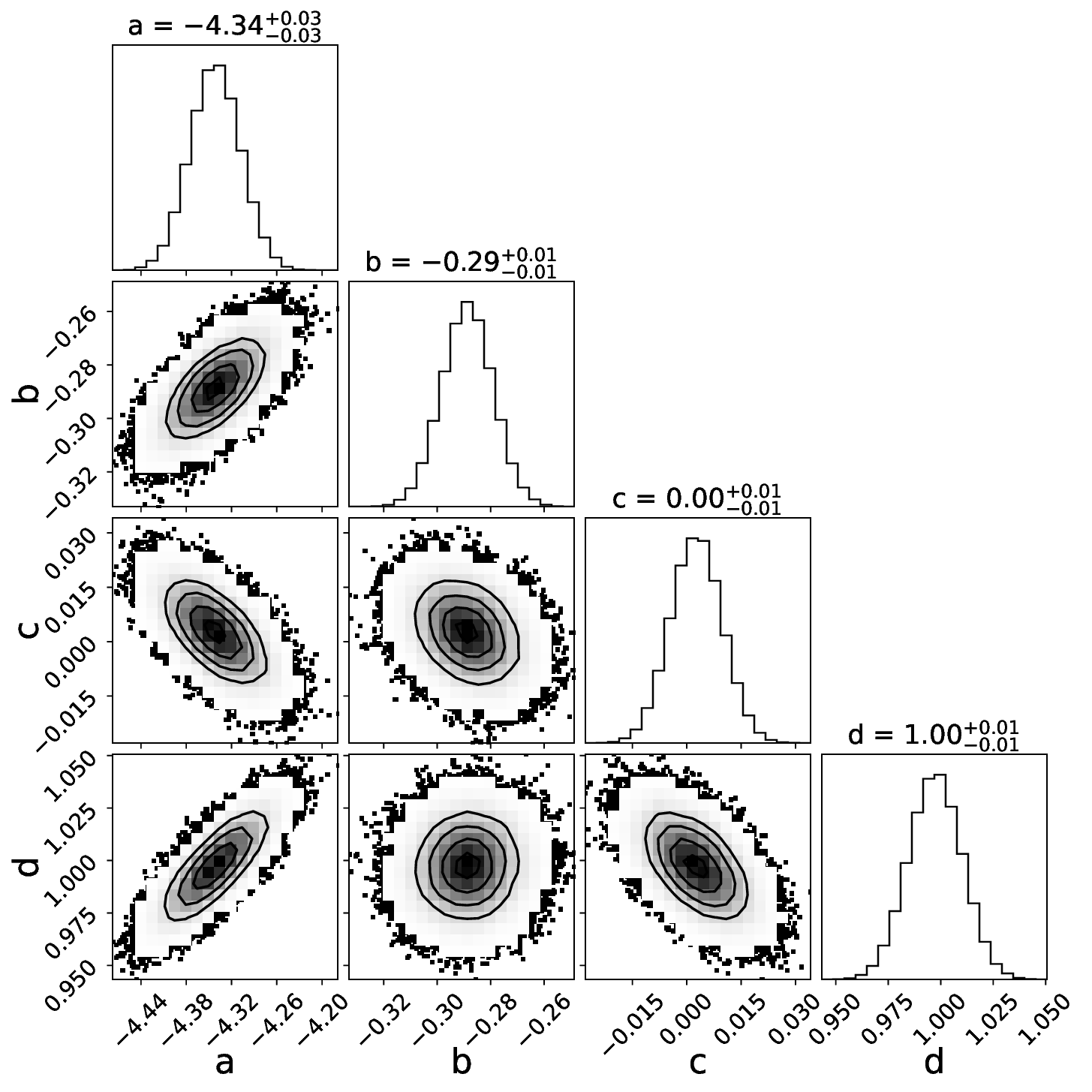}{0.325\textwidth}{(7) W2-band}\hspace{-15mm} 
          \fig{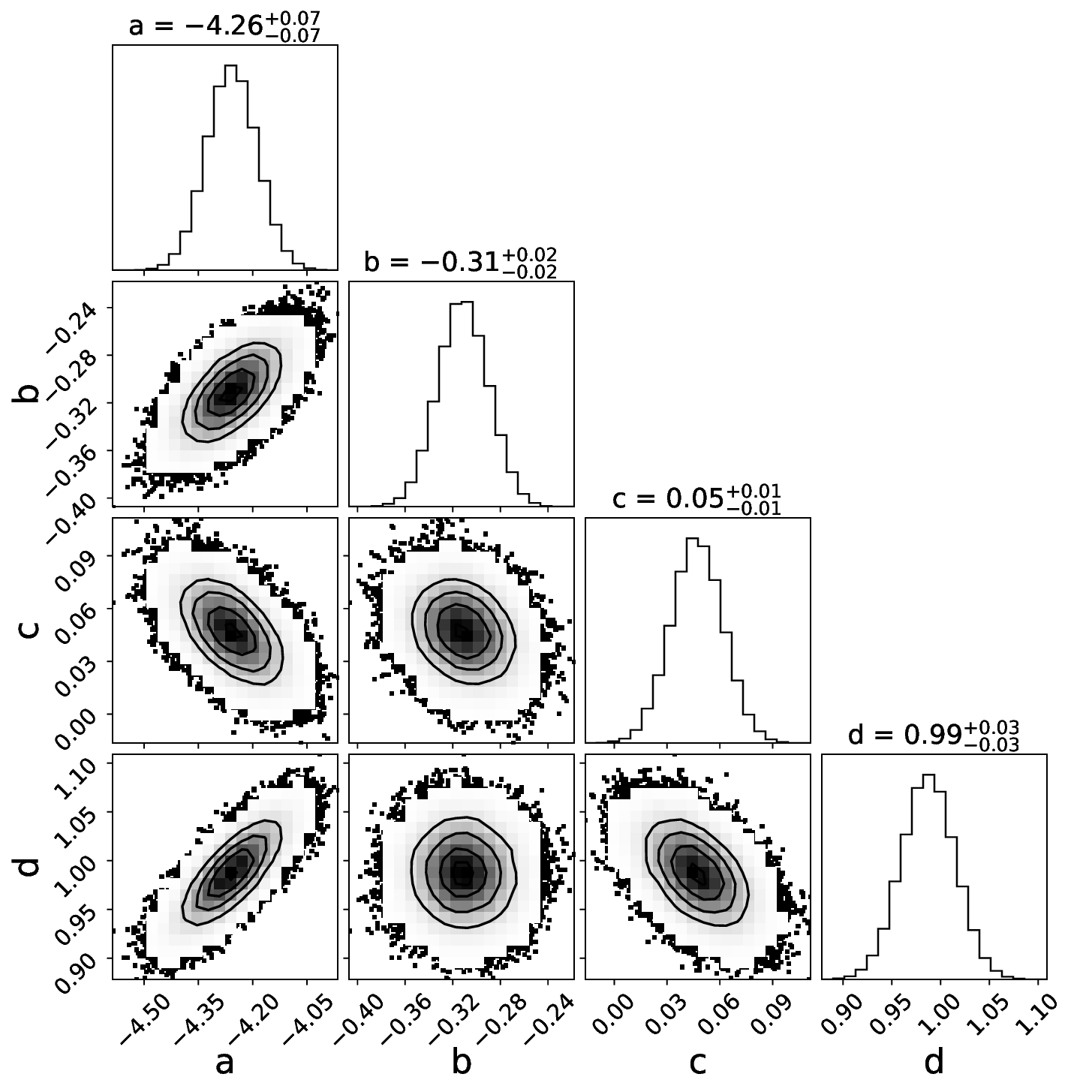}{0.325\textwidth}{(8) W3-band}\hspace{-15mm} 
          \fig{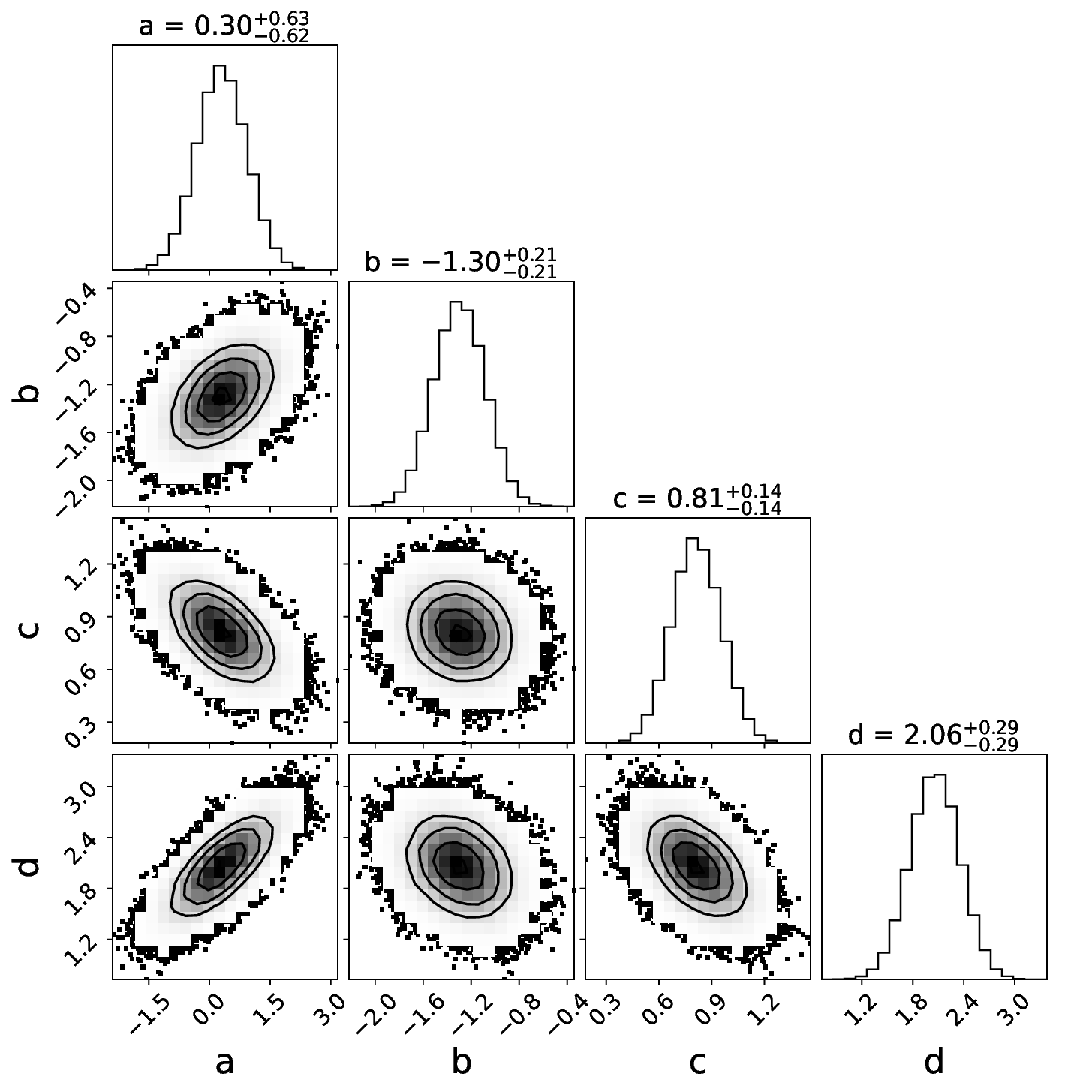}{0.325\textwidth}{(9) W4-band}}
\caption{Parameters fitting for PLZC relations by {\tt emcee}.
\label{fig.9}}
\end{figure*}

Then, we consider the influence of these two terms on PL relations at the same time, 
that is, PLZC relations. 
The performances of fitting PLZC relations based on equation \ref{equ4} are shown in Fig. \ref{fig.9}. 
Table \ref{table5} summarizes their fitting parameters and dispersion.
The coefficients of $\rm [Fe/H]$ in Table \ref{table5} are 
much smaller than these of color index, which demonstrates that 
calibrations could also make sense if metallicity term is not available 
when color information is provided. 
\begin{deluxetable}{ccccccch}
\setlength{\tabcolsep}{5mm}
\tablecaption{Parameters of PLZC models for multi-band. \label{table5}}
\tablewidth{5pt}
\tablehead{Band & Number &  a & b & c & d & $\sigma$ (mag)}
\startdata
$FUV$ & 199  & $-$8.32($\pm0.42$) & 7.24($\pm0.14$) & 1.21($\pm0.10$) &3.01($\pm0.18$)& 0.59  \\
$NUV$ & 632  & $-$7.29($\pm0.05$) & 0.30($\pm0.02$)  & 1.20($\pm0.01$)& 6.51($\pm0.02$)&0.31 \\
\hline
$B_{p}$ &1,093& $-$5.24($\pm0.03$) & $-$0.71($\pm0.01$) & 0.06($\pm0.01$)&3.04($\pm0.01$) &0.18 \\
$G$  & 1,093 & $-$5.65($\pm0.02$) & $-$0.66($\pm0.01$) & 0.11($\pm0.01$) &2.40($\pm0.01$)& 0.17 \\
$R_{p}$& 1,093 &$-$5.25($\pm0.03$) & $-$0.71($\pm0.01$) & 0.06($\pm0.01$)& 2.04($\pm0.01$)&0.18\\
\hline
$W1$ & 1,019 & $-$4.28($\pm0.03$) & $-$0.29($\pm0.01$) & $-$0.02($\pm0.01$)& 0.99($\pm0.01$)&0.12  \\
$W2$ & 1,019  & $-$4.34($\pm0.03$) & $-$0.29($\pm0.01$) & 0.00($\pm0.01$) & 1.00($\pm0.01$)&0.12\\
$W3$ & 992  & $-$4.26($\pm0.07$) & $-$0.31($\pm0.02$) & 0.05($\pm0.01$)&0.99($\pm0.03$)& 0.22   \\
$W4$  & 127  & 0.30($\pm0.63$) & $-$1.30($\pm0.21$)  & 0.81($\pm0.14$) & 2.06($\pm0.29$)&1.15  \\
\enddata
\tablecomments{The parameters a, b, c, and d are best-fitting coefficients in equation \ref{equ4}, 
and $\sigma$ are the deviations of PLZC relations.}
\end{deluxetable}
\subsection{Calibrations for Synthetic Photometry\label{synthetic}}
In addition to calibrations to the measured data, relations based on synthetic photometry are also determined here. 
The Gaia Synthetic Photometry Catalogue ($\rm GSPC$) is provided by \citet[]{2022arXiv220606215G},
utilizing the GaiaXPy\footnote{https://gaia-dpci.github.io/GaiaXPy-website/} Python package.
In any suitable system, synthetic photometry by XP spectra (XP Synthetic Photometry, XPSP, hereafter) can provide optical photometry for billions of stars. 
When synthetic photometry is standardized in relation to these external sources, it can achieve millimag accuracy for wide and medium bands.
As discussed in the methods of \citet[]{2022arXiv220606215G}, synthetic photometry is based on the 
convolving a properly normalised mean flux \citep[as defined in][]{2012PASP..124..140B} with a transmission curve.
Therefore, it is equally applicable to binary systems with mean flux measured.
Here we use the standardised Johnson-Kron-Cousin’s \citep[$\rm JKC$,][]{2022A&A...664A.109P}, $\rm SDSS$ \citep[]{2021MNRAS.505.5941T}, 
PanSTARRS-1 \citep[$\rm PS1$,][]{2020ApJS..251....6M,2022AJ....163..185X} and Hubble Space Telescope \citep[$\rm HST$,][]{2018MNRAS.481.3382N} 
bands provided in the GSPC to calibrate PLC and PLZC relations. As shown by \citet[][]{2022arXiv220606215G}, the summary of GSPC content for each 
passband is listed in their Table 5.
\begin{deluxetable}{cccccch}
\setlength{\tabcolsep}{5mm}
\tablecaption{Parameters of PLC models for synthetic photometry. \label{table6}}
\tablewidth{5pt}
\tablehead{Band & Number &  a & b & d & $\sigma$ (mag)}
\startdata
$U_{\rm JKC}$ & 207 & $-$5.36($\pm0.15$)  & $-$0.80($\pm0.06$)& 8.53($\pm0.12$)&0.22  \\
$B_{\rm JKC}$ & 865 & $-$5.96($\pm0.03$)  & $-$0.23($\pm0.01$) & 5.80($\pm0.02$)&0.27\\
$V_{\rm JKC}$  & 883 & $-$5.93($\pm0.02$)  & $-$0.31($\pm0.01$)&4.34($\pm0.01$)& 0.24   \\
$R_{\rm JKC}$  & 874 & $-$5.59($\pm0.02$)   & $-$0.31($\pm0.01$) & 3.83($\pm0.01$)&0.19  \\
$I_{\rm JKC}$   & 879  & $-$5.27($\pm0.02$)  & $-$0.31($\pm0.01$)& 3.42($\pm0.01$)&0.15 \\
\hline
$u_{\rm SDSS}$ &245 & $-$5.44($\pm0.14$) & $-$0.02($\pm0.05$)& 8.47($\pm0.11$)&0.23  \\
$g_{\rm SDSS}$ & 867& $-$5.92($\pm0.02$) & $-$0.39($\pm0.01$) & 5.27($\pm0.01$)&0.26\\
$r_{\rm SDSS}$  & 839 & $-$5.68($\pm0.02$) & $-$0.18($\pm0.01$)&3.89($\pm0.01$)& 0.20  \\
$i_{\rm SDSS}$  & 874 & $-$5.39($\pm0.02$)   & $-$0.02($\pm0.01$) & 3.54($\pm0.01$)&0.16  \\
$z_{\rm SDSS}$   & 877 & $-$5.03($\pm0.02$)  & 0.19($\pm0.01$)& 3.33($\pm0.01$)&0.15  \\
\hline
$y_{\rm ps1}$ &774 & $-$4.74($\pm0.04$) & 0.35($\pm0.01$)& 3.37($\pm0.02$)&0.14  \\
\hline
$F606W_{\rm ACS/WFC}$ & 863 & $-$5.27($\pm0.40$)  & $-$0.03($\pm0.08$) & 2.66($\pm0.17$)&0.20\\
$F814W_{\rm ACS/WFC}$  & 882 & $-$4.94($\pm0.03$) & $-$0.09($\pm0.01$)&2.16($\pm0.01$)& 0.16   \\
\enddata
\end{deluxetable}

The number of sources used for calibration in each band are displayed in Tables 
\ref{table6} and \ref{table7}, and the best-fitting parameters and dispersion 
of calibrations are also included in the tables. The dispersion of PLZC relations is reduced 
to values between 0.12 and 0.20 mag when adopting $\rm XPSP$ for calibrations. 
Comparing to the results based on measured data in Sect. \ref{uv} and these from \citet{CDD+2018}, 
the dispersion of PLC or PLZC relations based on $\rm XPSP$ has a significant improvements.
The improvements benefit probably from that $\rm XPSP$ are generated from the 
Gaia $B_{P}/R_{P}$ mean spectra and standardised using wide and reliable sets of external standard stars.
Part of the reason comes from the GaiaXPy, which is used to analyze XP spectra, 
by default provides an underestimated nominal uncertainty for the synthetic 
flux \citep{2022arXiv220606143D}, and the measured data typically contains some inevitable and significant systematic errors. 
\begin{deluxetable}{ccccccch}
\setlength{\tabcolsep}{3mm}
\tablecaption{Parameters of PLZC models for synthetic photometry. \label{table7}}
\tablewidth{5pt}
\tablehead{Band & Number &  a & b & c & d & $\sigma$ (mag)}
\startdata
$U_{\rm JKC}$ & 207 & $-$5.59($\pm0.19$) & $-$0.83($\pm0.06$) & 0.09($\pm0.03$)& 8.35($\pm0.02$)&0.22  \\
$B_{\rm JKC}$ & 865 & $-$6.13($\pm0.03$) & $-$0.26($\pm0.01$) & 0.06($\pm0.01$) & 5.69($\pm0.01$)&0.27\\
$V_{\rm JKC}$  & 883 & $-$6.10($\pm0.03$) & $-$0.34($\pm0.01$) & $-$0.06($\pm0.01$)&4.25($\pm0.02$)& 0.24   \\
$R_{\rm JKC}$  & 874 & $-$5.42($\pm0.02$) & $-$0.28($\pm0.01$)  & $-$0.06($\pm0.01$) & 3.93($\pm0.01$)&0.19  \\
$I_{\rm JKC}$   & 879  & $-$4.81($\pm0.03$) & $-$0.21($\pm0.01$) & $-$0.15($\pm0.01$)& 3.69($\pm0.02$)&0.15 \\
\hline
$u_{\rm SDSS}$ &245 & $-$5.57($\pm0.19$) & $-$0.04($\pm0.05$) & 0.05($\pm0.04$)& 8.37($\pm0.09$)&0.23  \\
$g_{\rm SDSS}$ & 867& $-$6.01($\pm0.03$) & $-$0.40($\pm0.01$) & 0.04($\pm0.01$) & 5.21($\pm0.02$)&0.26\\
$r_{\rm SDSS}$  & 839 & $-$5.46($\pm0.03$) & $-$0.13($\pm0.01$) & $-$0.07($\pm0.01$)&4.02($\pm0.01$)& 0.20  \\
$i_{\rm SDSS}$  & 874 & $-$4.95($\pm0.03$) & $-$0.07($\pm0.01$)  & $-$0.14($\pm0.01$) & 3.80($\pm0.02$)&0.16  \\
$z_{\rm SDSS}$   & 877 & $-$4.42($\pm0.03$) & 0.32($\pm0.01$) & $-$0.19($\pm0.01$)& 3.69($\pm0.02$)&0.15  \\
\hline
$y_{\rm ps1}$ &774 & $-$4.09($\pm0.05$) & 0.47($\pm0.01$) & $-$0.21($\pm0.01$)& 3.75($\pm0.03$)&0.14  \\
\hline
$F606W_{\rm ACS/WFC}$ & 863 & $-$5.80($\pm0.02$) & $-$0.29($\pm0.01$) & $-$0.05($\pm0.01$) &4.07($\pm0.01$)&0.20\\
$F814W_{\rm ACS/WFC}$  & 882 & $-$4.79($\pm0.03$) & $-$0.22($\pm0.01$) & $-$0.16($\pm0.01$)&3.71($\pm0.02$)& 0.16   \\
\enddata
\end{deluxetable}

\subsection{Testing and applying the PLZC relations\label{distance tracers}}
Considering the PLZC relations are more accurate than the other relations for estimating magnitudes, 
here we adopt the calibrated PLZC relations to estimate the distances of CBs. 
Since the PLZC relation of the $W1$-band has the smallest dispersion among these bands,
we utilized it by converting equation \ref{amag} to estimate distances $D_{CB}$, i.e.,
\begin{equation}
\label{d}
\log D_{CB}={\frac{m-M_{model}-A_{\lambda} +5 }{5}}, 
\end{equation}
where $M_{model}$ is the absolute magnitude derived through equation \ref{equ4}. 

\begin{figure}[ht!]
\centering\includegraphics[width=0.5\columnwidth]{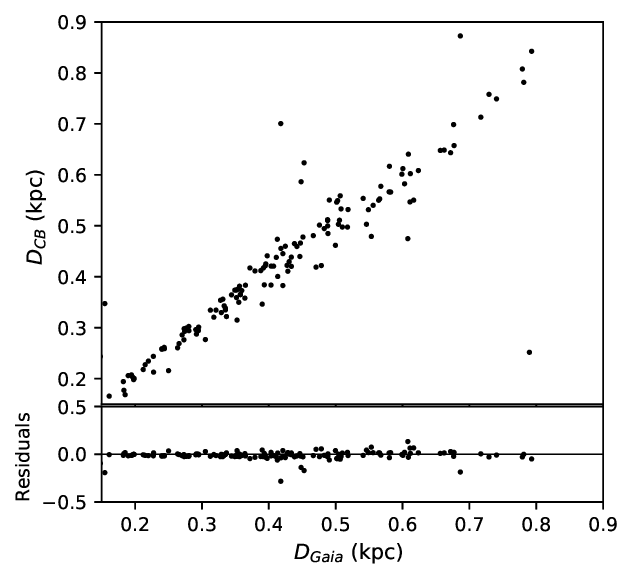}
\caption{Comparisons of the distances derived from the PLZC relations and these from Gaia DR3. The residuals of distances is presented in below the sub-figure. \label{fig10}}
\end{figure}

A catalog of 388 independent CBs are obtained from \citet[]{2015A&A...578A.136L}, 
from where we select 148 CBs containing spectroscopic and photometric information as test samples.
The distances of the 148 CBs are determined based on equation \ref{d}. 
Comparisons of the derived distance with these from Gaia DR3 are shown in Fig. \ref{fig10}. 
The accuracy of derived distance is estimated to be 5\% (0.11 mag for $\delta$) through the independent test.
Moreover, we test the distances of 2 CBs (CRTS J0254 and J0121) derived from the calibrated PLZC relations with high-precision distances from \citet {2022RAA....22i5017M}. 
In this work, the distances of the systems are measured as 1017 and 1247 pc, respectively. The estimations are in good agreement with 1023 and 1087 pc from \citet {2022RAA....22i5017M} with deviations of 1\% and 15\%. 

Stars in open clusters (OCs) have roughly the same age, distance, and chemical composition, while independent distances of OCs can be measured precisely.
Most OCs located at high latitudes in the Milky Way are less effected by the extinction, which are preferred tools for testing calibrations. 
We derive the distances of CBs which are recognized as OCs membership or candidates. 
The name of the OCs are listed in the first column of Table \ref{table8}, followed with their distances in the second column 
which are estimated through analyzing cluster memberships based on Gaia DR3 by \citet[]{2023A&A...673A.114H}.
The $\rm RA$, $\rm Dec$ and periods of CBs as OCs membership (or candidates) are given in the 3-5th columns, while the corresponding literatures in the seventh column. 
The derived distances of the CBs ($D_{CB}$) are listed in the sixth column. 
The $\delta D $ in the last column presents the difference between $D_{OC}$ and $D_{CB}$ ($\delta D = D_{OC} - D_{CB}$). 
For the majority of the CBs, who are OC members, a comparison of the $D_{CB}$ and $D_{OC}$ reveals good agreement.
For a certain CBs, the large discrepancies between $D_{CB}$ and $D_{OC}$ imply that those CBs are probably foreground or background sources of the OCs. 
For example, the two CBs in NGC 957 are more likely to be foreground stars in the cluster, 
which is consistent with the study of \citet[]{2017NewA...52...29L}.
\begin{deluxetable}{cccccccch}
\setlength{\tabcolsep}{1.5mm}
\tablecaption{Distance comparisons of CBs and OCs. \label{table8}}
\tablewidth{5pt}
\tablehead{Cluster &  $D_{OC}$&  RA    & Dec   &  Period      & \textbf{$D_{CB}$}& Ref. of membership or candidates & \textbf{$ \delta D $ }\\
                   &    (pc)  &  (deg) & (deg) &  (day)       & (pc)   &                                  &  (pc)              }
\startdata
NGC 957   & 2384 $\pm$ 620   & 38.595  & 57.429   &  0.40     & 872$\pm$46   & \citet[]{2009OEJV..112....1B} & 1512 $\pm$666 \\
          &                  & 38.108  & 57.445   &  0.31     & 674$\pm$84   &                               & 1710 $\pm$704\\
\hline     
NGC 1912  & 1140 $\pm$ 581   & 82.059  & 35.591   &  0.38     & 1541$\pm$19  & \citet[]{2021RAA....21...68L} & $-$401 $\pm$600\\
          &                  & 81.964  & 35.722   &  0.46     & 1028$\pm$84  &                               &112 $\pm$665\\
          &                  & 82.299  & 35.722   &  0.38     & 1375$\pm$40  &                               &$-$235 $\pm$621\\
\hline     
NGC 1245  & 3362$\pm$ 586    & 48.617  & 47.185   &  0.40     & 3308$\pm$10  & \citet[]{2006AJ....132.1177P} &54 $\pm$596\\
\hline
King 7    & 3324 $\pm$ 282   & 60.270  & 51.361   &  0.36     & 698$\pm$59   & \citet[]{2008IBVS.5857....1B} & 2626 $\pm$ 341\\
          &                  & 60.200  & 51.196   &  0.29      & 2030$\pm$3  &                              &1294 $\pm$285\\
\hline   
NGC 2126  & 1316 $\pm$ 287   & 90.895  & 49.612   &  0.44     & 1922$\pm$15  & \citet[]{2009RAA.....9..791L}& $-$606 $\pm$302\\
\hline   
NGC 2168  & 867  $\pm$ 679   & 92.371  & 24.117   &   0.39    & 2074$\pm$2  & \citet[]{2005ChJAA...5..356H}&$-$1207 $\pm$681\\
          &                  & 91.779  & 24.401   &   0.32    & 1798$\pm$10  &                               &$-$931 $\pm$689\\
\hline   
NGC 2355  & 1903 $\pm$ 370  & 109.349 & 13.781   &   0.31    & 1404$\pm$33  & \citet[]{2022AJ....164...40W}&499 $\pm$403\\
\hline   
NGC 6819  & 2661$\pm$ 1115   & 296.165 & 40.080   &   0.30    & 1108$\pm$97  & \citet[]{2021AJ....161...35L}&1553 $\pm$1212\\
          &                  & 294.929 & 40.147   &   0.44    & 387$\pm$33   &                               &2274 $\pm$1148\\
          &                  & 295.196 & 40.350   &   0.46    & 2789$\pm$28  &                               &$-$128 $\pm$1143\\
          &                  & 295.224 & 40.526   &   0.42    & 2035$\pm$44  &                               &626 $\pm$1159\\
\hline   
NGC 6939  & 1905 $\pm$ 889   & 308.341 & 60.620   &   0.29    & 1732$\pm$23  & \citet[]{2008AN....329..387M}&173 $\pm$912\\
\hline   
NGC 7142  & 2437 $\pm$ 522   & 326.118  & 65.776  &  0.33     & 270$\pm$14   & \citet[]{2011PZP....11...32P}&2167 $\pm$536\\
\hline   
NGC 7789  & 2087$\pm$ 1404   & 359.455  & 56.783  &  0.33     & 2086$\pm$25  & \citet[]{1995AA...295..101J}&1 $\pm$1429\\
\enddata
\end{deluxetable}
\section{Discussions and Conclusions} \label{sec:dis_con}
Considering relations between color index and stellar luminosity, as well as the influences of metallicity abundance on stellar radius and luminosity, 
PLZC relations for late-type CBs are calibrated based on current surveys and synthetic photometry. Starting with collecting CBs from photometry surveys and catalogs, 
we cross-match them with LAMOST and Gaia catalogs to obtain the spectroscopic data 
and distance information to construct a sample of 1093 CBs to calibrate. 
For the sample, PLZC relations for optical, infrared, and ultraviolet bands are derived. 
As comparisons, PL, PLZ, and PLC relations are also derived simultaneously. 
In addition to these relations for measured data, relations based on synthetic photometry are also calibrated. 

Based on the PLZC relations of CBs for infrared and optical bands, distances could be measured with an accuracy of 6\% (0.12 mag) and 8\% (0.17 mag), respectively. 
In the optical bands, the dispersion of PLZC relations could be reduced to 6\% (0.14 mag) when using the CBs sample with GSPC. 
In addition, the fitting results of FUV, NUV, W3, and W4-band are relatively poor, probably due to the small sampling rates of the observations. 
GALEX visits its targets with at least one epoch, mostly more than twice \citep{BS+2017}.
W1 and W2-band have been observed with more than 30 epochs, while W3 and W4-band visited with $\sim$16 epochs \citep{2010AJ....140.1868W}. Gaia has made at least 15-epoch observations for its each band \citep{2023A&A...674A...1G}.

Pollutions from other variables to CBs would affect the estimation of distances. Though variable stars such as RR Lyrae with periods ranging from 0.2 to 1.0 day have the similar periods to CBs, RR Lyrae with redder colors (than CBs) are easy to identify from CBs. As discussed in \citet{2014ApJS..213....9D}, contamination (from RRc) leads to larger CB distances on the order of 1\% (0.02 mag), which is lower than the uncertainty of calibrations in this work. Thus, pollutions from RR Lyrae are ignored here.
A certain proportion of close binaries are accompanied by tertiary companions \citep{1982Ap&SS..88..433R,1998ApJ...504..978H}, which is an important source of systemic uncertainty for calibrations and distance estimations. 
According to \citet{2006AJ....132..650D},
the uncertainty in total luminosity introduced by a relatively bright tertiary is less than 15\% (or 0.15 mag).
The calibration uncertainties caused by tertiary companions could be reduced to $0.15/\sqrt{N}$ mag, where $N$ is the number of samples for calibration. 
Its typical value would be less than 0.01 mag for a sample with hundreds of CBs. 
The systemic uncertainty of distance estimation introduced by a faint tertiary star would be less than $\sim$7\% even smaller. 
It is also expected to be effectively reduced by increasing samples. 
Since the sample for calibrations in this work are not tested for tertiary companions, 
the influence of tertiary companions is included in the calibrations by default. 

Tests of the distances derived from PLZC relations with these from Gaia and literatures demonstrate the robust of the calibrations. 
Comparing the distances of CBs with the average distance of OCs could reveal whether the CBs are OCs memberships or not. 
Given the periods are well determined, the uncertainty of distances estimated based on the method depend on the uncertainties of photometry and extinction. Assuming the limiting magnitude of a photometry survey is about 24 mag \citep[e.g. ESA’s Euclid mission’s step$-$and$-$stare survey strategy covers the full area to a depth of YJH $\sim$ 24 mag,][]{2016ASPC..507..401J}, the PLZC method could measure the CBs as far as 60 Kpc, which covers the distance of LMC \citep[e.g.][]{CDD+2016}.

Although considering the influence of [$\rm Fe/H$] and color index could improve the precision of model calibrations, 
it probably remains some improvements if the intrinsic luminosity are well studied from an theoretical view. 
In future work, we are going to analyze the Period-Luminosity-Color and Period-Luminosiy-Metallicity-Color relations based on stellar models and quantify the relations from a theoretical view. 

\begin{acknowledgments}

This work has made use of products from the Guoshoujing Telescope (the Large Sky Area Multi-Object Fibre Spectroscopic Telescope, LAMOST). LAMOST is a National Major Scientific Project by the Chinese Academy of Sciences. Funding for the project has provided by the National Development and Reform Commission. LAMOST is operated and managed by the National Astronomical Observatories,Chinese Academy of Sciences.

This work is supported by National Key R$\&$D Program of China No. 2019YFA0405503 and 2022YFF0503404, 
National Natural Science Foundation of China 11803030, 11803029, 12090040, 12090042, Joint Research Fund in Astronomy U203120019, 
Science \& Technology Department of Yunnan Province - Yunnan University Joint Funding (2019FY003005), 
Fundamental research project of Yunnan Province 202101AT070012, and the science research grants from the China Manned Space Project (No. CMS-CSST-2021-A10).

\end{acknowledgments}

\bibliography{sample631}{}
\bibliographystyle{aasjournal}



\end{document}